\documentclass{aa}  
\usepackage{graphicx}
\usepackage[figuresright]{rotating}
\usepackage{natbib}
\bibpunct{(}{)}{;}{a}{}{,}
\usepackage{savesym}
\usepackage{amsmath}
\savesymbol{iint}
\usepackage{txfonts}
\restoresymbol{TXF}{iint}
\usepackage{graphicx}   
\usepackage{verbatim}   
\usepackage{color}      
\usepackage{subfigure}  
\begin{document}

   \title{Polycyclic aromatic hydrocarbon processing by cosmic rays}

   \author{E. R. Micelotta\inst{1,2,3,4}, A. P. Jones\inst{2}, A. G. G. M. Tielens\inst{1,5}}

   \offprints{E. R. Micelotta}

   \institute{Sterrewacht Leiden, Leiden University, P.O. Box 9513, 2300 RA 
              Leiden, The Netherlands\\
              \email{elisabetta.micelotta@nasa.gov}
              \and
              Institut d'Astrophysique Spatiale, Universit\'{e} Paris Sud and CNRS (UMR 8617),
              91405 Orsay, France
              \and
              CRESST and NASA Goddard Space Flight Center, Greenbelt, MD 20771, USA 
              \and
              Universities Space Research Association, 10211 Wincopin Circle, 
              Suite 500 Columbia, MD 21044, USA 
              \and
              NASA Ames Research Center, MS 245-3, Moffett Field, CA 94035, USA\\ 
             }

   \date{Received 13 September 2010; accepted 22 October 2010}

  \abstract
   {Cosmic rays are present in almost all phases of the ISM. 
    PAHs and cosmic rays represent an abundant and ubiquitous
    component of the interstellar medium. However, the interaction
    between them 
    has never before been fully investigated.}
   {To study the effects of cosmic ray ion (H, He, CNO and Fe-Co-Ni) 
    and electron bombardment of PAHs in galactic and extragalactic
    environments.}
   {We calculate the nuclear and electronic interactions for
    collisions between PAHs and cosmic ray ions and electrons with
    energies between 5 MeV/nucleon and 10 GeV, above the threshold for carbon
    atom loss, in normal galaxies, starburst galaxies and cooling
    flow galaxy clusters.}
   {The timescale for PAH destruction by cosmic ray ions
    depends on the electronic excitation energy $E_0$ and on the
    amount of energy available for dissociation. Small PAHs are
    destroyed faster, with He and the CNO group being the more
    effective projectiles.
    For electron collisions, the lifetime is independent of the PAH
    size and varies with the threshold energy $T_0$.}
    {Cosmic rays process the PAHs in diffuse clouds, where the
     destruction due to interstellar shocks is less efficient.  In
     the hot gas filling galactic halos, outflows of starburst
     galaxies and intra-cluster medium, PAH destruction is dominated
     by collisions with thermal ions and electrons, but this
     mechanism is ineffective if the molecules are in denser
     cloudlets and isolated from the hot gas. Cosmic rays can access
     the denser clouds and together with X-rays will set the lifetime
     of those protected PAHs. This limits the use of PAHs as a
     `dye' for tracing the presence of cold entrained material.
    }

   \keywords{dust, extinction -- cosmic rays -- Galaxies: halos -- 
             Galaxies: starburst -- ISM: jets and outflows -- shock waves -- 
             Galaxies: cooling flows -- intergalactic medium
}
   \authorrunning{E. R. Micelotta et al.}

   \titlerunning{PAH processing by cosmic rays}
   \maketitle


\section{Introduction}

A remarkable characteristic of 
the IR emission features is that they dominate the mid-IR spectrum of almost all
objects associated with dust and gas and illuminated by UV photons, ranging 
from {\sc H\,ii} regions to ultraluminous infrared galaxies 
\citep[see][ for a recent review]{tielens08}.
At present, these feature
are (almost) universally attributed to the IR fluorescence of
far-ultraviolet (FUV)-pumped Polycyclic Aromatic Hydrocarbon (PAH)
molecules containing 50 -- 100 carbon atoms \citep{tielens08}.
PAHs have recently been detected in association with shocked hot gas,
but it is difficult to establish a clear connection between the two.
\citet{tappe06} detected spectral features in the emission of the 
supernova remnant \object{N132D} in the Large Magellanic Cloud, which they attribute 
to emission by large PAHs. 
\citet{reach06} have identified four supernova
remnant with IR colors maybe indicating PAH emission, and \citet{and07}
investigated the presence of PAHs in a subset of galactic supernova
remnants in the GLIMPSE survey. PAHs have also been observed interwoven
with the X-ray emission arising from the bipolar outflow of the starburst
galaxy M82 \citep{armus07} and in the high-latitude coronal gas of the
edge-on galaxies \object{NGC 5907}, \object{NGC 5529} and \object{NGC 891}
\citep{irwin06, irwin07,
whaley09}.

Unfortunately the lack of theoretical studies on PAH processing in
shocked regions combined with the difficulty in disentangling the 
PAH features intrinsic to the shocked region with those arising from 
the surrounding material makes the interpretation of such observations 
rather complicated. In our previous works \citep[][ hereafter MJTa and MJTb]
{micelotta09b, micelotta09a}  
we studied the 
survival of aromatic molecules in interstellar shocks with velocities 
between 50 and 200 km s$^{-1}$ and in a hot post-shock gas, such as
the Herbig- Haro jets in the Orion and Vela star forming regions
\citep{podio06}, in the local interstellar cloud \citep{slavin08} and
in the outflow of the
starburst galaxy M82 \citep{engel06}.  We found that interstellar PAHs
($N_{\rm C} = 50$) do not survive in shocks with velocities greater
than 100 km s$^{-1}$ and larger PAHs ($N_{\rm C} = 200$) are destroyed
for shocks with velocities $\geq 125$ km s$^{-1}$. Even where
destruction is not complete, the PAH structure is likely to be
severely denatured by the loss of an important fraction ($20-40$\%) of
the carbon atoms.  The typical PAH lifetimes are of the order of a few
$\times 10^8$ yr for the Galaxy.  In a tenuous hot gas ($n_{\rm H}
\approx 0.01$ cm$^{-3}$, $T \approx$ 10$^7$ K), typical of the coronal
gas in galactic outflows, PAHs are principally destroyed by electron
collisions, with lifetimes measured in thousands of years, i.e. orders of magnitude
shorter than the typical lifetime of such objects.  

Cosmic rays (CRs) are an important component of the ISM, contributing considerably 
to its energy density
\citep[$\simeq$ 2 eV cm$^{-3}$, ][]{tielens05, padovani09}. CRs consist mainly of
relativistic protons, $\alpha$-particles ($\sim$ 10\%), and heavier
ions and electrons ($\sim$ 1\%). 
The spectrum (intensity as a function of the energy) of the ionic CR 
component measured near the Earth spans from $\sim$ 100 MeV to $\sim$
10$^{20}$ eV, and decreases steeply with energy. The
spectrum of the electronic component is even steeper and ranges from
$\sim$ 600 MeV to 10$^3$ GeV \citep{ip85, gaisser06}.

The lowest-energy CRs in the ISM, with energy between 5 MeV and
few GeV, are excluded from the heliosphere or severely slowed down
by the solar wind. Hence, they cannot be directly observed even with
far-ranging spacecraft \citep{shapiro96} and their spectra have to be
evaluated theoretically \citep{shapiro91}.

CRs with energy up the few 10$^{15}$ eV \citep[the ``knee'' 
observed in the spectrum: ][]{drury94} are thought to be produced
in the Galaxy, mainly by supernova shocks in the disk. Because
of their charge, CRs are tied to the galactic magnetic field and are
confined to a spheroidal volume with radius of $\sim$ 20 kpc and
half-thickness of $\sim$ 1 -- 15 kpc \citep{ginzburg88, shibata07},
with a small but finite escape probability. The magnetic field
randomizes the trajectories of CRs as they propagate through the Galaxy,
so their distribution is almost isotropic except close to the sources.

From the point of view of PAH destruction, CRs have then two
interesting characteristics: first, for energies up to 10 GeV they can
efficiently transfer energy to the PAH, with possible consequent
destruction (see Sects.\ 2 and 3); second, they permeate almost
homogeneously the ISM and can penetrate into regions such as dense
clouds which are otherwise not much affected by high temperature ions
and electrons (MJTb).

The aim of this work is to quantify the destructive potential of
CRs and to compare it with other mechanisms (interstellar
shocks, collisions within a hot gas, X-ray and FUV absorption), in
galactic and extragalactic environment.

The paper is organized as follows:
Sect. 2 and Sect. 3 describe the treatment of high energy ion and electron 
interactions with PAHs, Sect. 4 presents the CR spectra adopted for
our study and Sect. 5 illustrates the calculation of the collision rate
between PAHs and CRs. We present our results on PAH destruction 
and lifetime in Sect. 6 and discuss the astrophysical implications
in Sect. 7, summarizing our conclusions in~Sect.~8.

\section{High energy ion interactions with solids}

\subsection{Collisions with high energy ions}

To describe the effects of high-energy ion collisions with PAH
molecules we adopt a similar approach to that used in our previous
work (MJTa and MJTb), based on the theory of ion interaction in
solids. Ions colliding with a PAH will excite the molecule and the 
excitation can lead to fragmentation or, alternatively, the excess
energy can be radiated away. A calculation of the fragmentation process
thus consists of two steps: 1) the calculation of the excitation
energy after collision, 2) the probability of dissociation of an
excited PAH. The former is discussed here. The latter is described in
section Sect.\ 2.2.

Energetic ions can interact with a solid through
two simultaneous processes 
\citep{lind63}: 
the \textit{nuclear stopping} or \textit{elastic energy loss}, where the
energy is directly transferred from the projectile ion to a target nucleus 
via a binary elastic collision, and the \textit{electronic stopping} or 
\textit{inelastic energy loss}, consisting of the energy loss to the electron
plasma of the medium. 
For high energies characteristic of CRs (above $\sim$ 1 MeV/nucleon) 
we only need to be concerned by
electronic stopping, which can be 
well described by the Bethe--Bloch equation \citep[see e.g.]
[ and references therein]{ziegler99}.

The Bethe--Bloch equation was derived considering the electromagnetic
interaction of an energetic particle with the electron plasma of a solid.
A PAH molecule is not a solid but its large number of delocalized electrons
can be treated as an electron gas (see MJTb and references therein). It is 
therefore appropriate to consider the energy loss to such an electron plasma in 
terms of the Bethe--Bloch electronic stopping power, $S$ (energy loss per unit 
length). When $S$ is known, a specific procedure has to be applied to calculate 
the effective amount of energy transferred into a single molecule, taking into
account the finite geometry of the PAH (see Sect.\ 3).

The stopping of high velocity ions in matter has been a subject of interest
for more than a century, starting with the work of Marie Curie in 1898-1899
\citep{mcurie1900}. For a theoretical review on the topic, we refer the reader
to \citet{ziegler99} and references therein. In the following, we summarize the
basic methods and equations for evaluating the stopping power of high energy
ions and we illustrate the modifications introduced into the theory in order 
to treat the interaction with PAH molecules.

For high energy light ions (H, He and Li above 1 MeV/nucleon), the fundamental relation 
describing the stopping power, $S$, in solids is the relativistic Bethe-Bloch 
equation, commonly expressed as
\begin{eqnarray}\label{S1_eq}
  S & = & \frac{\kappa Z_2}{\beta^2}\,Z_1^2\,L(\beta) \\
    & = & \frac{\kappa Z_2}{\beta^2}\,Z_1^2\,\left[L_0(\beta)+Z_1L_1(\beta)+
      Z_1^2L_2(\beta) ...\right]
\end{eqnarray}
where
\begin{eqnarray}\label{kappa_eq}
  \kappa \equiv 4\pi\,r_0^2\,m_{\rm e}c^2,\;\;\;
                r_0 \equiv e^2/m_{\rm e}c^2\;\;\rm{(Bohr\:electron\:radius)}
\end{eqnarray}
$m_{\rm e}$ is the electron mass, $Z_1$ and $Z_2$ are the projectile and 
target atomic numbers respectively and $\beta = v/c$ is the relative projectile
velocity.
The term $L(\beta)$ is defined as the stopping number, and its expansion
in Eq.\ 2  contains all the corrections to the basic ion-electron energy
loss process.

The first term $L_0$ includes the fundamental Bethe-Bloch relation 
\citep{bethe30, bethe32, bloch33a, bloch33b} for the stopping of high-energy 
ions, together with the two corrective terms $C/Z_2$ and $\delta/2$ introduced 
by \citet{fano63}
\begin{eqnarray}\label{L0_eq}
  L_0 = f(\beta)\,-\,\frac{C}{Z_2}\,-\,
      \ln \langle I \rangle\,-\,\frac{\delta}{2}
\end{eqnarray}
where
\begin{eqnarray}\label{f_eq}
  f(\beta) \equiv \ln\,\left[\frac{2\,m_{\rm e}c^2\,\beta^2}{1\,-\,\beta^2}\right]\,-\,\beta^2
\end{eqnarray}

$C/Z_2$ is the \textit{shell correction} and takes into account the fact that as
soon as the projectile loses energy into the target, its velocity decreases from 
relativistic values, thus the Bethe-Bloch theory requirement of having particles
with velocity far greater than the velocity of the bound electron is no longer
satisfied. In this case, a detailed accounting of the projectile's interaction with
each electronic orbital is required. The shell correction is important for protons
in the energy range of 1 -- 100 MeV, with a maximum contribution of about 10\%. 

$\ln \langle I \rangle$, which is part of the original Bethe-Bloch relativistic stopping
formula, represents the \textit{mean ionization}, and corrects for the fact that 
the energy levels of the target electrons are quantized and not continuous.

$\delta/2$ represents the \textit{density effect} and provides the correction to 
the reduction of the stopping power due to polarization effects in the target. 
The dielectric polarization of the target material reduces the particle
electromagnetic field from its free-space values, resulting in a variation of the
energy loss. The density effect becomes important only for particles with kinetic
energies exceeding their rest mass (938 MeV for proton).

The term $Z_1\,L_1$ takes into account the Barkas effect. This is due to the target
electrons, which respond to the approaching particles slightly changing their orbits
before the energy loss interaction can occur.

The term $Z_{\rm 1}^{\rm 2}L_2$ is the Bloch correction and provides the transition
between the two approaches used to evaluate the energy loss of high-energy 
particles to target electrons: the classical Bohr impact--parameter approach 
\citep{bohr13, bohr15}, and the quantum-mechanical Bethe momentum transfer approach
\citep{bethe30, bethe32}.  

Both the Barkas and Bloch corrections are usually quite small and contribute
less than few percent to the stopping at energies from 1 MeV/nucleon to 10$^4$ MeV.

The term $L_0$ can be evaluated as a function of the energy of the incoming ion
using the following definition \citep{ziegler99}:
\begin{eqnarray}\label{beta2Ion_eq}
  \beta^2 \equiv \left(\frac{v}{c}\right)^2\;=\;1\,-\,\frac{1}
       {\left[1\,+\,E(\rm{GeV})\,/\,0.931494\;\it{M}_{\rm 1} (\rm{amu})\right]^2}
\end{eqnarray}
Using Eq. \ref{L0_eq}, we can rewrite the expression for the stopping power (Eq.\ 2)
in the following way
\begin{multline}\label{stoppingTotal_eq}
  S = \frac{\kappa\,Z_2}{\beta^2}\,Z_1^2\,\Bigg\lbrace\,\left[f(\beta)\,-\,\frac{C}{Z_2}\,-\,
      \ln \langle I \rangle\,-\,\frac{\delta}{2}\right]\,+\,   \\
     Z_1\,L_1(\,\rm {Barkas}\,)\,+\,\it{Z}_{\rm 1}^{\rm 2}L_2(\,\rm {Bloch}\,)\Bigg\rbrace 
\end{multline}

For the stopping power in units of eV/(10$^{15}$ atoms/cm$^2$) the
prefactor constant has the value $\kappa$ = 5.099$\times$10$^{-4}$, while
for the stopping units of eV/\AA\,the above prefactor has to be multiplied
by $N$/10$^{23}$, where $N$ is the atomic density of the target (atoms/cm$^3$).
In our calculation we adopt the values for amorphous carbon ($Z_2$ = 6, $M_2$ = 12, 
$N$ = 1.1296$\times$10$^{23}$ atoms/cm$^3$), which yields $\kappa$ = 5.7508$\times$10$^{-4}$.

An empirical expression for the Barkas correction term is available \citep{ziegler99}, 
whereas the Bloch correction can be evaluated using the Bichsel parametrisation 
\citep{bichsel90}. \citet{bond67} 
estimates that $\langle I \rangle = 11.4\,Z_2$ (eV), but
this is not always in agreement with experimental data and unfortunately there are 
no simple algorithms for the shell correction and density effect. As a consequence,
no simple analytical expressions for the stopping power are available.

The SRIM program \citep{zie85} calculates accurate stopping powers
from Eq. \ref{stoppingTotal_eq} using different methods to evaluate
the corrective terms.  The shell correction $C/Z_2$ is the
average of the values obtained from the Local Density Approximation
theory (LDA) and Hydrogenic Wave Function (HWF) approach
\citep{ziegler99}. The first is an \textit{ab initio} calculation
based on realistic solid state charge distributions, while the second
uses parameterized functions based on experimental stopping data.
For the density correction $\delta/2$, the values tabulated in \citet{icru84}
are used, while the term $\ln \langle I \rangle$ is derived by adjusting the 
theoretical value obtained from the LDA theory \citep[e.g.][]{lind52} in order
to fit the sum $\ln \langle I \rangle + C/Z_2$ evaluated empirically from
experimental stopping data.

As already mentioned, Eq. \ref{stoppingTotal_eq} is valid for light ions, H, He 
and Li. The stopping power for ions with $Z_1 > $ 3 is usually calculated using
the heavy-ion scaling rule, as reported for example by \citet{katz72}
\begin{eqnarray}\label{heavy_eq}
  S(Z_1,\,\beta) = S(p,\,\beta)[Z_1^*/Z_p^*]^2
\end{eqnarray}
where $S(p,\,\beta)$ is the stopping power of a proton at the same speed as the ion
of atomic number $Z_1$ (Eq. \ref{stoppingTotal_eq}), $Z_1^*$ and $Z_p^*$ are the 
effective charge numbers of ion $Z_1$ and of a proton respectively, with the expression 
given by Barkas \citep[see e.g.][]{hen70}. 
\begin{eqnarray}\label{zeff_eq}
  Z_1^* = Z_1\,\left[1\,-\,\exp\left(-125\,\beta\,Z_1^{-2/3}\right)\right]
\end{eqnarray}

To verify the effective importance of the corrective terms to the Bethe-Bloch
equation in our specific case of interest (H, He, C and Fe impacting on carbon),
we compared the output from SRIM with the stopping power calculated from the 
following approximate equation
\begin{eqnarray}\label{stoppingApprox_eq}
  S = \frac{\kappa\,Z_2}{\beta^2}\,Z_1^2\,\left[f(\beta)\,-\,
      \ln \langle I \rangle\right]
\end{eqnarray}
where we adopt the carbon mean ionization energy calculated by 
SRIM\footnote{http://www.srim.org/SRIM/SRIMPICS/IONIZ.htm} 
using the method described above, $\langle I \rangle$ = 79.1 eV. For the heavier 
ions C and Fe we use the proton stopping power from Eq. \ref{stoppingApprox_eq} 
into Eq. \ref{heavy_eq}.

\begin{figure}
  \centering
 \includegraphics[width=1.\hsize]{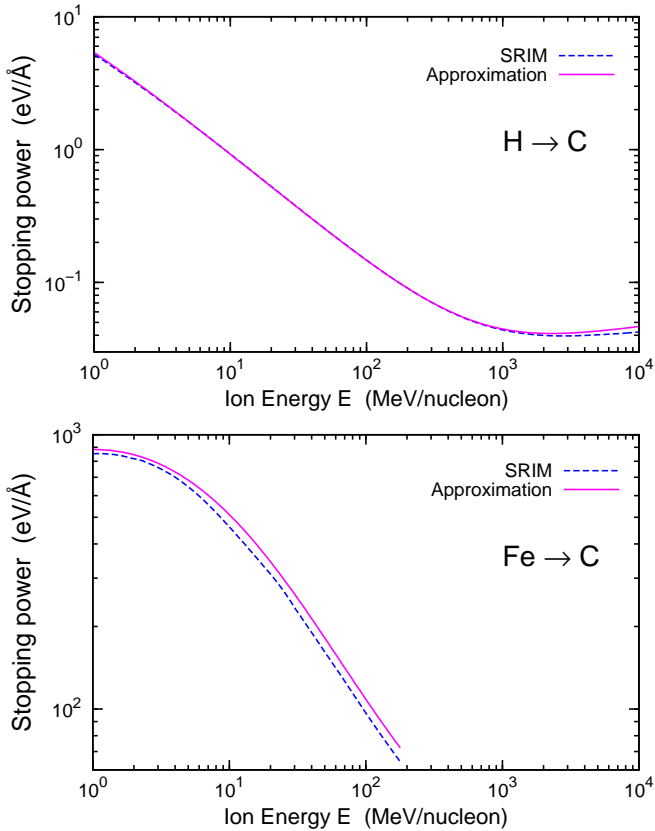}
  \caption{Stopping power of hydrogen and iron impacting on carbon, 
           as a function of the energy per nucleon of the ion.
           The lower validity limit of the
           Bethe-Bloch equation is 1 MeV/nucleon. We compare
           the output from the SRIM code, which computes all
           corrections, and our approximate equation which includes
           only the mean ionization correction.
           }
  \label{ionStopping_fig}
\end{figure}

We find a maximum discrepancy between the two curves of $\sim$10\%,
very small with respect to the other astrophysical uncertainties and 
indicating that our approximate equation is adequate to describe the
energy loss. In this specific case, not only the Barkas and Bloch
corrections, but also the shell correction and the density effect play
a marginal role. Fig. \ref{ionStopping_fig} shows the comparison
between the `accurate' SRIM curve and the approximate one for H and
Fe, our lightest and heaviest projectiles respectively. 

Using Eq. \ref{heavy_eq}  and \ref{stoppingApprox_eq} we can then calculate
the electronic stopping power d$E$/d$x$ of energetic ions and from this
the energy loss to the PAH molecule (Sect.\ 3). Since the stopping
power is a decreasing function of the ion energy and increases quadratically
with $Z_1$, the major contribution to the energy loss will come from the less 
energetic particles for a given ion, and from the heavier species for
a given velocity (Fig. \ref{ionStopping_fig}).


\subsection{Ion energy loss and dissociation probability}

\begin{figure}
  \centering
\includegraphics[width=1.\hsize]{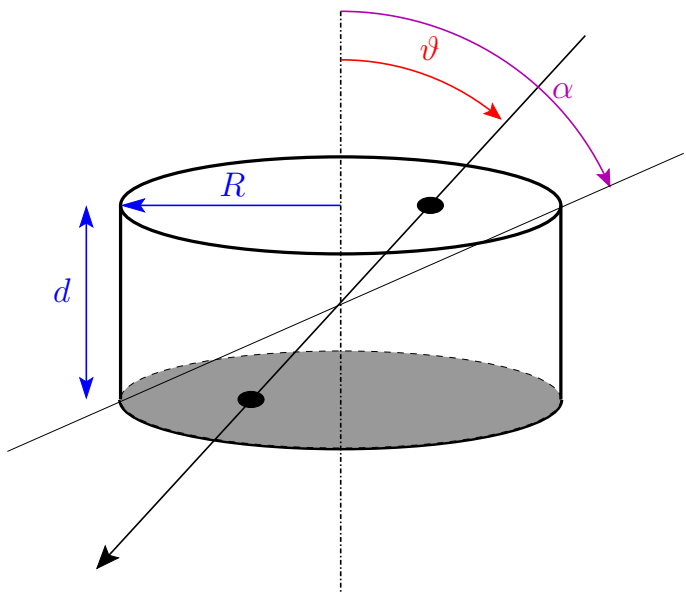}
  \caption{Coordinate system adopted to calculate the energy transferred
           to a PAH via electronic excitation by ion collisions. The molecule
           is modeled as a disk with radius $R$ and thickness $d$. The trajectory
           of the incoming particle is identified by the angle $\vartheta$, while
           the angle $\alpha$ corresponds to the diagonal of the disk.
           }
  \label{ionConfig_fig}
\end{figure}

To calculate the energy transferred to a PAH
during collisions with high energy ions we adopt the
configuration shown in Fig.~\ref{ionConfig_fig} (see
also MJTb).


The molecule is modeled as a thick disk with radius $R$ given by 
the usual expression for the radius of a PAH: 
$a_{\rm PAH} = R = 0.9\,\sqrt(N_{\rm C})\;$\AA, where $N_{\rm C}$ is the number 
of carbon atoms in the molecule \citep{omont86}. For a 50 C-atom PAH, 
$R = 6.36\,$\AA. The thickness of the disk, $d\sim\:$4.31\,\AA\, is the thickness
of the electron density assumed for the PAH (see MJTb).

The path $l$, through the PAH, along which the incoming ion loses its energy is
defined by the impact angle $\vartheta$ and by the dimensions of the
molecule. Inspection of Fig. \ref{ionConfig_fig} shows that, if
$|\tan(\vartheta)| < \tan(\alpha)$, $l(\vartheta) =
d/|\cos\vartheta|$, otherwise $l(\vartheta) = 2R/|\sin\vartheta|$. The
rigorous method to calculate the energy loss along $l(\vartheta)$
takes into account the progressive slowing down of the projectile in
traversing the target.  This is explained in Sect. 3 of MJTb. 
Nevertheless, in the present case we consider high energy
particles for which the energy loss along the path $l$ is small
compared to the initial energy (cf. Fig. \ref{ionStopping_fig}). In
other words, the energy of the incoming ion remains almost constant
during the interaction, thus the amount of energy lost after
travelling the distance $l(\vartheta)$ can be simplified to
\begin{eqnarray}\label{eTransfIon_eq}
  T(\vartheta) = \frac{{\rm d}E}{{\rm d}x}\times l(\vartheta)
\end{eqnarray}
where d$E/$d$x$ = $S(\vartheta)$ is the stopping power (energy lost
per unit length) from
Eq. \ref{stoppingApprox_eq} and/or Eq. \ref{heavy_eq}, and d$E$ is 
the amount of energy lost after travelling the distance d$x$.
The difference
between the results from the two methods is of the order of few
percent, which fully justifies the use of approximation 
in Eq.~\ref{eTransfIon_eq}.

The ion collision will leave the molecule electronically excited. 
Internal conversion and/or intersystem crossing will transfer this 
excitation (largely) to the vibrational manifold. De-excitation can 
occur through two competing decay channels:
emission of infrared
photons and dissociation and loss of a C$_{2}$ fragment.  The latter
is the process that we are interested in because it leads to PAH
fragmentation. 
The emission of a C$_{2}$ fragment is suggested by
experiments on fullerene C$_{60}$ which have shown that the ejection
of C$_{2}$ groups is one of the preferred fragmentation channel. 
Moreover, the loss of acetylene groups C$_{2}$H$_{2}$ has
been observed in small PAHs. In fact, in a PAH molecule, a side group
C$_{2}$H$_{n}$ (with $n$ = 0, 1, 2) is easier to remove because only two
single bonds have to be broken, while the ejection of a single
external C-atom requires one single and one double bond to be broken,
and for an inner C-atom from the skeleton three bonds needs to be
broken.
To quantify the PAH destruction due to ion collisions
we need to determine the probability of dissociation, $P$, rather than
IR emission. For the detailed calculation and a discussion
of the dissociation probability, we refer the reader to Sect. 4.1 in
MJTb. For the sake of clarity, the basic equations are reproduced here.
The total dissociation probability is calculated by combining the rates 
for fragmentation ($k_0 \exp[-E_0/k\,T_{\rm av}]$) and IR decay 
($k_{\rm IR}/(n_{\rm max}+1)$) into the following expression
\begin{eqnarray}\label{totalProb2_eq}
  P = \frac{k_{\rm 0}\,\exp \left[-E_{\rm 0}/k\,T_{\rm av} \right]}
         {\left[k_{\rm IR}/(n_{\rm max}+1)\right]\:+\:k_{\rm 0}\,
          \exp \left[-E_{\rm 0}/k\,T_{\rm av} \right]}
\end{eqnarray}
where $k_0$ and $E_0$ are the Arrhenius pre-exponential factor and 
energy describing the fragmentation process respectively, $k_{\rm IR}$ 
and $n_{\rm max}$ are the IR photon emission rate and number of IR photons 
(MJTb), and $k$ is the Boltzmann constant.
The temperature $T_{\rm av}$ is chosen as the geometrical mean between
two specific effective temperatures of the PAH
\begin{eqnarray}\label{Tav_eq}
  T_{\rm av} = \sqrt{T_{\rm 0}\times T_{n_{\rm max}}}
\end{eqnarray} 
In the microcanonical description of a PAH,
the temperature, $T$, describing the excitation (for fragmentation purposes) 
is related to the internal energy, $E'$, by
\begin{eqnarray}\label{teff1_eq}
  T \simeq 2000\,\left(\frac{E'(\rm eV)}{N_{\rm C}}\right)^{0.4}\,
                    \left(1\,-\,0.2\,\frac{E_{\rm 0}(\rm eV)}{E'(\rm eV)}\right)
\end{eqnarray}
where $E_{\rm 0}$ is the binding energy of the fragment
\citep{tielens05}. 
The temperatures $T_0$ and $T_{n_{\rm max}}$ in Eq. \ref{Tav_eq} are the 
temperatures when the internal energy equals the initial transferred energy 
($E'= E$) and when the internal energy equals the energy after emission of 
$n_{\rm max}$ photons ($E'= E - n_{\rm max} \times \Delta \varepsilon$) 
with $\Delta \varepsilon$ being the average energy of the emitted IR photon.
For the number of photons, $n_{\rm max}$, required to be emitted to have 
the probability per step drop by an order of magnitude, we adopt 10, the 
average photon energy is set equal to 0.16 eV, corresponding to a typical 
CC mode, and the pre--exponential is set equal to
1.4~$\times$~10$^{16}$~s$^{-1}$
 (MJTb). For the photon emission rate we adopt the typical value 
$k_{\rm IR}$ = 100 photons s$^{-1}$ \citep{joc94}.

From the above equations one can see that $P$ depends on the binding
energy of the fragment, $E_{0}$, on the PAH size, $N_{\rm C}$, and on
the energy transferred, $T$, which in turns depends on the initial
energy of the projectile. For a fixed value of the transferred energy,
the dissociation probability decreases for increasing $E_{0}$ and
$N_{\rm C}$ because either more energy is required in the bond that
has to be broken or because the energy is spread over more vibrational
modes and hence the internal excitation temperature is lower. On the
other hand, the more energy that is deposited in the PAH, the higher
is the dissociation probability.

The fragment binding energy $E_{0}$, which is a crucial parameter in
the evaluation of the dissociation probability is, unfortunately, presently 
not well constrained. As in our previous work (MJTb) we investigated the
impact on the PAH destruction process for $E_{0}$ = 
3.65, 5.6 and 4.58 eV, the latter value is consistent with extrapolations 
to interstellar conditions and is our reference value.

As mentioned at the end of Sect. 2.1, the stopping power increases with the
atomic number of the projectile and decreases with its energy. Thus, for a 
given pathlength, the energy transferred, and therefore the dissociation 
probability, will be higher for low energy heavy particles.

\section{Collisions with high energy electrons}

To model the interaction of high--energy electrons with PAH molecules, 
we refer to the formalism used to describe the irradiation effects in
solid materials, in particular carbon nanostructures \citep{ban99}.
In the collisions of high-energy (relativistic) electrons with nuclei,
the screening effect of the surrounding electrons is negligible. The
electron--nucleus interaction can thus be treated in terms of a binary
collision using a simple Coulomb potential, applying the appropriate
relativistic corrections \citep[e.g.][]{reimer89, ban99}.

If the energy transferred to the nucleus exceeds the \textit{displacement
energy} $T_{\rm d}$, i.e. the minimum energy required to produce a vacancy--
interstitial pair which does not spontaneously recombine, the atom will be 
knocked out. If its energy is above the threshold value for further
displacements, it can remove other atoms in its environment generating a
collision cascade.

In this description the ``bulk'' nature of the target enters only after the
first interaction, when projectile and displaced atom propagate into the 
solid. Therefore, if we limit ourselves to the first interaction only, 
this approach can be applied to electron--PAH collisions and allows us to 
take into account the ``molecular'' nature of our target.
In fact, this is the same binary collision approach used to describe the 
\textit{nuclear interaction} (elastic energy loss) in collisions between 
PAHs and relatively low energy ions in interstellar shocks (MJTa).

\begin{figure}
  \centering
 \includegraphics[width=.9\hsize]{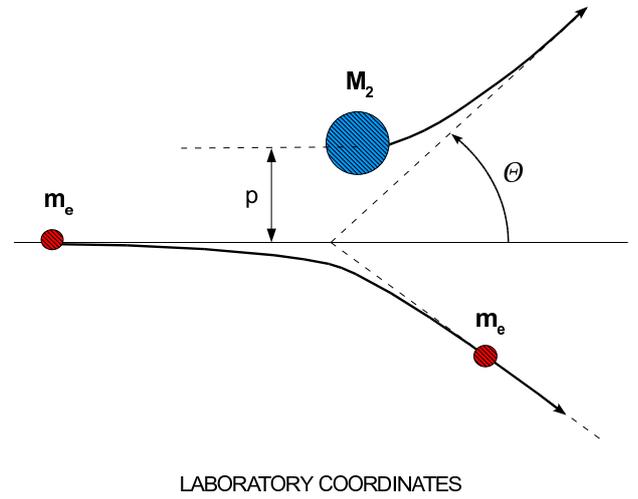}
  \caption{
           Scattering geometry for the elastic collision of an
           electron (mass $m_{\rm e}$, impact parameter $p$, initial
           velocity $v$) on a massive particle (mass
           $M_{2}$, initial velocity zero). After the impact, the
           target particle is knocked out and recoils at an angle
           $\Theta$ with respect of the initial direction of motion of
           the projectile electron.
           }
  \label{electronScattering_fig}
\end{figure}

A PAH is a planar molecule with tens to hundreds of carbon atom. In this case the 
target nucleus is a single carbon in the PAH. If the energy transferred exceeds 
a threshold value the target 
nucleus will be ejected from the molecule. The displacement energy then has to be 
replaced by the threshold energy $T_{\rm 0}$, which represents the minimum
energy to be transferred in order to knock-out a carbon atom from a PAH. 
The interaction between a high energy CR electron and a PAH
occurs between the impinging electron and one single target carbon
atom in the molecule. The energy is transferred from the projectile
electron to a target carbon atom through a binary collision. Thus,
each electron collision implies the loss of one single carbon atom, 
in contrast to ion collisions where each interaction causes the ejection
of a C$_2$ fragment from the PAH molecule.
The scattering geometry is shown in Fig. \ref{electronScattering_fig}.
After the collision, the target nucleus is knocked out and recoils
at an angle $\Theta$ with respect of the initial direction of 
motion of the projectile electron. The energy $T$ transferred to the
nucleus depends on the scattering angle
\begin{eqnarray}\label{T_eq}
  T(\Theta) = T_{\rm max}\cos^2\Theta
\end{eqnarray}
The term $T_{\rm max}$ is the maximum transferable energy, corresponding 
to a head--on collision ($\Theta$ = 0), and is given by the following 
equation \citep{simmons65}  
\begin{eqnarray}\label{TmaxElec_eq}
  T_{\rm max} = \frac{2\,E\,\left(E\,+\,2\,m_{\rm e}c^2\right)}
              {M_2\,c^2}
\end{eqnarray}
where $E$ is the electron kinetic energy, $m_{\rm e}$ is the electron mass and
$M_2$ is the target atomic mass.


The total displacement/removal cross section, $\sigma$, i.e. the cross
section for collisions able to transfer more than the threshold energy
$T_{\rm 0}$, is defined as the integral over the solid angle of the
differential cross section d$\sigma$/d$\Omega$, which provides the
probability for atomic recoil into the solid angle d$\Omega$
\begin{eqnarray}\label{sigmaElectronTOT_eq}
  \sigma = \int_{0}^{\Theta_{\rm
      max}}\frac{{\rm d}\sigma}{{\rm d}\Omega}\,2\pi\,\sin\Theta\,{\rm d}\Theta
\end{eqnarray}
where $\Theta = 0$ corresponds to the transfer of $T_{\rm max}$ and
$\Theta = \Theta_{\rm max}$ is the recoil angle corresponding to the
transfer of the minimun energy $T_{\rm 0}$. The calculation of the
total cross section $\sigma$ for atom displacement/removal would
require the analytical treatment of the Mott scattering of a
relativistic electron by a nucleus \citep{mott29, mott32}. The
corresponding equations have to be solved numerically, but
\citet{mckin48} found an analytical approximation which provides
reliable values of $\sigma$ for light target elements such as carbon
(under the assumption of an isotropic displacement/threshold
energy). We adopt the formulation of the analytical expression
reported by \citet{ban99}
\begin{multline}\label{sigmaElectron_eq}
  \sigma  =  \frac{4\,Z_2^2\,E_{\rm R}^2}{m_{\rm e}^2\,c^4}\;\left(\frac{T_{\rm max}}{T_{\rm 0}}\right)\,
               \pi\,a_{\rm 0}^2\;\left(\frac{1\,-\,\beta^2}{\beta^4}\right)\;  \\
               \left\{1\,+\,2\,\pi\,\alpha\,\beta\,\,\left(\frac{T_{\rm 0}}{T_{\rm max}}\right)^{1/2}\right.  \\
\left.          -\,\frac{T_{\rm 0}}{T_{\rm max}}\;\left[1\,+\,2\,\pi\,\alpha\,\beta\,+\,
               \left(\beta^2\,+\,\pi\,\alpha\,\beta \right)\,\ln\,\left(\frac{T_{\rm max}}{T_{\rm 0}}\right)\,\right]\,\right\}
\end{multline}
where $Z_2$ is the atomic number of the displaced atom (in our case,
carbon), $E_{\rm R}$ is the Rydberg energy (13.6 eV), $m_{\rm e}c^2$ is
the electron rest mass (0.511 MeV), $a_{\rm 0}$ is the Bohr radius of
the hydrogen atom (5.3$\times$10$^{-11}$ m), $\beta = v/c$, $v$ being
the velocity of the incident electron, and $\alpha$ = $Z_2$/137, where
1/137 is the fine structure constant.

The term $T_{\rm 0}$ is the minimum energy which has to be transferred
into the PAH in order to remove a carbon atom and represents the
analog of the displacement energy $T_{\rm d}$ in a solid. For an
extended discussion about the determination of the threshold energy
$T_{\rm 0}$ we refer the reader to MJTb (Sect. 2.2.1). We recall
here that the value of $T_{\rm 0}$ is, unfortunately, not well
established, because there are no experimental determinations on PAHs
and the theoretical evaluation is uncertain. We decided to
explore possible values: 4.5 and 7.5 eV, close to the energy of
the single and double C-bond respectively and 15 eV, compatible with
the expected threshold for a single walled nanotube. We adopt 7.5 eV
as our reference value, consistent with all the experimental data.

\begin{figure}
  \centering
  \includegraphics[width=1.\hsize]{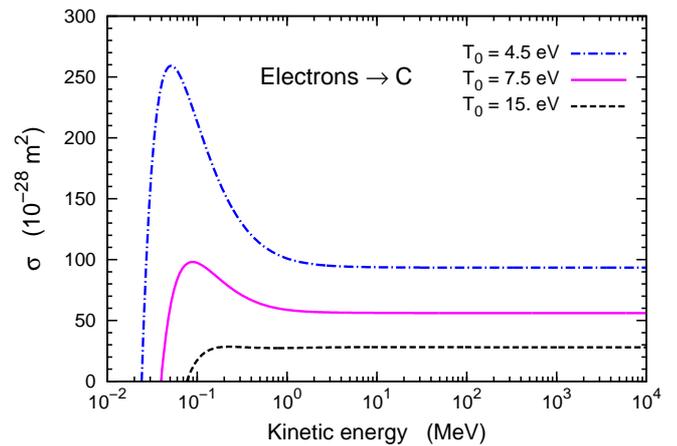}
  \caption{High energy electron cross section for carbon atom removal, 
           calculated for three values of the threshold energy $T_{\rm 0}$.
           }
  \label{eCrossSection_fig}
\end{figure}

To calculate $\beta$ as a function of the kinetic energy of the
incident electron, it is important to remember that we are considering
relativistic particles, thus the appropriate expression for the
kinetic energy is the following:
\begin{eqnarray}\label{EkinRel_eq}
  E_{\rm kin}^{\rm rel} = m_{\rm
    e}c^2\,\left(\frac{1}{\sqrt{1\,-\,\beta^{ 2}}}\,- 1\right)
\end{eqnarray} 
From Eq. \ref{EkinRel_eq} we then derive $\beta$
\begin{eqnarray}\label{beta_eq}
  \beta = \sqrt{1\,- \left(\frac{m_{\rm e}\,c^2}{E_{\rm kin}^{\rm
        rel}\,+\,m_{\rm e}\,c^2}\right)^2}
\end{eqnarray}
The displacement cross section $\sigma$ calculated from
Eqs. \ref{sigmaElectron_eq} and \ref{beta_eq}, is shown in
Fig. \ref{eCrossSection_fig} as a function of the electron kinetic
energy, for three different values of $T_{\rm 0}$. Above threshold the
cross section increases with electron energy and
decreases again at higher energies because of relativistic effects,
reaching the constant value given by the following asymptotic
expression
\begin{eqnarray}\label{sigmaAsym_eq}
  \sigma \sim \frac{8\,Z_2^2\,E_{\rm R}^2\,\pi\,a_{\rm 0}^2}{M_{\rm 2}\,c^2}\,\frac{1}{T_{\rm 0}}
\end{eqnarray}

As expected $\sigma$
decreases for increasing values of the threshold energy.  Around the
peak the change of $T_{\rm 0}$ from 4.5 to 15 eV introduces a
variation in the cross section of a factor of about 9, which reduces to 
3.4 - the ratio 15/4.5, cf. Eq. \ref{sigmaAsym_eq} -  for electron 
energies above $\sim$2 MeV.


\section{The CR spectrum}

The stopping power of ions with energy above $\sim$1~MeV/nucleon decreases 
with increasing energy (cf. Sect. 2.1). This implies that the CRs
responsible for the major energy transfer to a PAH, and subsequent
damage, are the lower-energy ones (below 1 GeV/nucleon). Unfortunately
this part of the CR spectrum is not accessible from the Earth because of the 
phenomenon called \textit{solar modulation} \citep{shapiro91}. CRs
entering the heliosphere see their intensity reduced by the effect
of the solar wind, especially at low energies and when the solar cycle
is at its maximum. The solar magnetic field is frozen within the plasma of 
the solar wind and drawn out with it into a spiral structure.
CRs encountering the solar wind are then convected
outward. Moreover when such charged particles interact with the 
expanding magnetic field, they are adiabatically decelerated. Hence, 
the CRs observed at a given energy were originally much more
energetic. 

Because of the solar modulation, the interstellar CR spectrum
at low energies needs to be evaluated theoretically by solving
the transport equation for particles in the ISM, assuming an
appropriate CR spectrum at the sources and taking into account all
possible mechanisms able to modify the intensity of the CRs
during their propagation \citep[energy losses, fragmentation etc. ]
[]{shapiro91}. 
The \textit{Pioneer} and \textit{Voyager} spacecraft
have probed the heliosphere out to beyond 60 AU, greatly improving
the understanding of the spectra of protons and heavier 
nuclei with energies above $\sim$ 100 MeV and the effects of solar modulation,
although limited information is available on CR nuclei below
$\sim$ 100 MeV \citep{webber98}.
To describe the propagation and escape of Galactic CRs
(at energies below a few $\times$10$^{15}$ eV), a widely used approach is the
leaky-box model, which assumes that the particles are confined to the
Galaxy, with frequent visits to the disk boundaries where they have
a small probability of leaking out \citep{ip85, simpson88, indriolo09}.

Following \citet{webber83} and \citet{bringa07} the CR intensity $I(E)$
as a function of the total energy per ion, $E$, is then given by
\begin{eqnarray}\label{ionSpectrum_eq}
  I(E) = C\,E^{0.3}\,/\,\left(E\,+\,E^*_0\right)^3\,\,\,
         \rm{\left(cm^2\;s\;sr\;GeV\right)^{-1}}
\end{eqnarray}
The constant $C$ can be determined by matching
Eq. \ref{ionSpectrum_eq} with the high-energy CR spectrum
measured on the Earth (see below). The scaling factor $E^*_0$ sets the
level of low-energy CRs \citep{webber83}, which is mainly
determined by ionization loss and Coulomb collisions \citep{ip85}. 
At higher energies (above $\sim$ 1 GeV/nucleon) diffusive losses 
dominate.

The interstellar CR spectrum can be constrained by molecular 
observations. CR protons ionize atomic and molecular hydrogen 
and this ionization drives interstellar chemistry through ion-molecule 
reactions. Analysis of molecular observations in diffuse clouds result 
in an average primary ionization rate of 4~$\times$~10$^{-16}$ s$^{-1}$ 
(H-nuclei)$^{-1}$ \citep[][ and references therein]{indriolo09}.
The CR ionization rate 
follows from a convolution of the CR spectrum, Eq. \ref{ionSpectrum_eq}, 
with the hydrogen ionization cross section \citep{bringa07}.
The scaling factor $E^*_0$ can be calculated then from
\begin{eqnarray}\label{ionizRate_eq}
  \zeta = 5.85\times10^{-16}\left(E^*_0 / 0.1\:\rm{GeV}\right)^{-2.56}
          \;\;\;\rm{s^{-1}\:\left(H\:nuclei\right)^{-1}}
\end{eqnarray}
Again, because of the steep decrease of the ionization  cross section with 
energy, the CR ionization rate is most sensitive to the low energy 
CR flux.
For hydrogen this results in $E^*_0\sim$~0.12 GeV. For heavier particles
we adopt the same scaling rule as \citet{bringa07}: 
$E^*_0$(ion)~=~$E^*_0$(H)$\times\,M_1$ GeV particle$^{-1}$, where $M_1$ is
the mass of the particle in amu.
 
To calculate the constant $C$, we matched Eq. \ref{ionSpectrum_eq} with
the high-energy spectrum detected on Earth. For the high energy data,
which are not influenced by the solar modulation, we adopt the
expression from \citet{wiebel98}
\begin{eqnarray}\label{ionHighSpectrum_eq}
  I(E) = I_0\,\left[E\,(\rm{GeV}) / (1000\,GeV)\right]^{-\gamma}
  \;\;\;\rm{\left(cm^2\;s\;sr\;GeV\right)^{-1}}
\end{eqnarray}
where $I_0$ and $\gamma$ depend on the CR ion.

\begin{figure}
  \centering
 \includegraphics[width=1.\hsize]{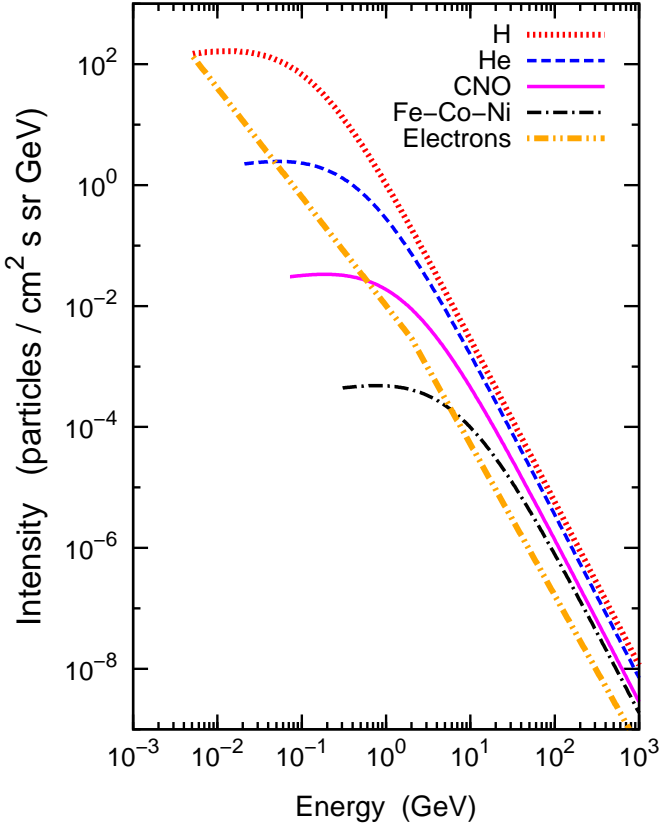}
  \caption{
           Interstellar CR spectrum of H, He, CNO, Fe-Co-Ni
           and electrons as a function of the particle energy.} 
  \label{CRspectra_fig}
\end{figure}

In this study we consider the most abundant CR components:
H, He, the group C, N, O and the group Fe, Co, Ni. The latter
are in fact often detected as a group because of the experimental difficulty
in distinguishing between particles with similar
mass. The spectra were calculated using the method described
above, with the high-energy parameters $I_0$ and $\gamma$ from 
\citet{wiebel98}. The matching between the low and high energy 
regimes is at $E$ = 1 TeV. A list of the parameters 
required for the calculation is reported in Table \ref{CRparams_tab}, 
and the resulting CR spectra are shown in Fig. \ref{CRspectra_fig}. 
For the lowest energy of the interstellar CR spectrum, we choose the
value of 5 MeV/nucleon, consistent with the limit of $\sim$ 1 MeV/nucleon adopted
by \citet{ip85} for their calculation of the CRs spectrum, and 
which corresponds to the
lower limit of the energy range where ionization loss rapidly diminishes the 
propagation of CRs in the ISM.


\begin{table}
  \begin{minipage}[t]{\columnwidth}
    \caption
        {CR ion spectra parameters.
        }
        \label{CRparams_tab}      
        \centering          
        \renewcommand{\footnoterule}{}      
        \begin{tabular}{c c c c c c}     
          \noalign{\smallskip}
          \noalign{\smallskip}
          \hline\hline       
          \noalign{\smallskip}
          Ion        &  $M_1^{(a)}$   &   $E^{*\,(b)}_0$   &  $C^{\,(c)}$    &  $I_0^{\,(d)}$  &  $\gamma$  \\  
          \noalign{\smallskip}
          \hline                    
          \noalign{\smallskip}         
          H          &   1.0  &   0.12     &  1.45   &  11.5$\times$10$^{-9}$  &  2.77     \\
          He         &   4.0  &   0.48     &  0.90   &  7.19$\times$10$^{-9}$  &  2.64     \\
          CNO        &   14.  &   1.68     &  0.36   &  2.86$\times$10$^{-9}$  &  2.67     \\
          Fe-Co-Ni   &   58.  &   6.95     &  0.24   &  1.89$\times$10$^{-9}$  &  2.60     \\    
          \noalign{\smallskip}
          \hline 
          \noalign{\smallskip}
        \end{tabular}        
  \end{minipage}
  ($a$): $M_1$ in amu. \\
  ($b$): $E^{*}_0$ (GeV) = $E^{*}_0$(H)$\times\,M_1$ (ion). \\
  ($c$): In units of (cm$^2$ s sr GeV$^{-1.7}$)$^{-1}$. \\ 
  ($d$): In units of (cm$^2$ s sr GeV)$^{-1}$. \\ 
\end{table}      


The same approach as that used for heavy particles (ions) can be applied to 
CR electrons. In this case solar modulation also alters
the spectrum of the electrons entering the solar cavity. The interstellar
spectrum at low energies has to be calculated solving the transport 
equation for electrons in the ISM, taking into account the energy
loss processes relevant for electrons, i.e. bremsstrahlung, synchrotron
and inverse Compton. The low energy spectrum then has to be connected to 
the measured high-energy spectrum, which is not affected by the modulation. 
We adopt the expression from \citet{cummings73}, calculated in the
framework of the leaky-box model \citep[see also][]{ip85, moska98}
\begin{eqnarray}\label{elec_eq}
  I(E) = A\,\left[E(\rm{GeV})\times10^3\right]^{-\gamma}\;\;\; \rm{\left(cm^2\;s\;sr \;GeV \right)^{-1}}
\end{eqnarray}
where $E$ is in GeV and \\

\smallskip
\noindent
$\begin{cases}
  \;A = 0.0254\times10^{5},\;\gamma = 1.8 & \text{for}\;5\times10^{-3} \leq E \leq 2, \\
  \;A = 5.19\times10^{5},\;\;\;\;\,\gamma = 2.5 & \text{for}\;2 < E \leq 10^{3}
\end{cases}
$\\

The calculated spectrum is shown in Fig. \ref{CRspectra_fig}. 
For the lowest energy, we again adopt a value of 5 MeV, coherent
with the range of influence of
bremsstrahlung, the dominant energy loss mechanism for electrons with energies
less than a few hundreds of MeV.

It is important to underline the fact that the method used to
calculate the interstellar CR spectrum at low energies, i.e. solving
the transport equation for particles adopting appropriate parameters
in order to match the measurements, does not provide an unique
solution.  It has been shown that different sets of parameters for the
transport model result in a set of different spectral shapes all
consistent with the CR intensity measured at higher energies
\citep[see e.g. ][]{mewaldt04}. Strong constraints on the low-energy
CR spectrum have been provided by the interstellar ionization rate
\citep[e.g. ][ and references therein]{indriolo09, padovani09},
which allowed to exclude some models, for instance the the W98 model 
\citep{webber98} for protons and the C00 model \citep{mosk02} for electrons.
Despite such improvement, the ambiguity on the low-energy region of the 
CR spectrum has not been completely solved yet. This uncertainty
does not affect the conclusions of our present work, but it is important
to keep it in mind.


\section{Collision rate and C-atom ejection rate}

To calculate the collision rate between PAHs and CR ions and
electrons, with energy fluxes described by Eq. \ref{ionSpectrum_eq} 
and \ref{elec_eq} respectively, we follow the procedure illustrated below.

For ion collisions, where the energy is transferred to the whole PAH via electronic
excitation, we first need to calculate the term $\Sigma(E)$, which takes into
account all possible ion trajectories through the PAH, with their corresponding
transferred energies and dissociation probabilities. Adopting the configuration shown 
in Fig. \ref{ionConfig_fig} we have
\begin{eqnarray}\label{SigmaTerm_eq}
  \Sigma(E) & = & \frac{1}{2\pi}\int {\rm d}\Omega\,\sigma_{\rm g}(\vartheta)\,P(E,\vartheta) \\
            & = & \int_{\vartheta=0}^{\pi/2}\sigma_{\rm g}(\vartheta)\,
                         P(E,\vartheta)\,\sin\vartheta\,{\rm d}\vartheta
\end{eqnarray}
with $\Omega = \sin \vartheta \,{\rm d}\vartheta\, {\rm d}\varphi$, with $\varphi$ running from 0 to 2$\pi$ 
and $\vartheta$ from 0 to $\pi/2$.
The geometrical cross section seen by an incident particle with direction defined by
the angle $\vartheta$ is given by
\begin{eqnarray}\label{disk_cs_eq}
  \sigma_{g}(\vartheta) = \pi\,R^{2}\,\cos\vartheta\,+\,2\,R\,d\,\sin\vartheta
\end{eqnarray}
which reduces to $\sigma_{g} = \pi \,R^{2}$ for $\vartheta$ = 0 (face-on
impact) and to $\sigma_{g} = 2\,R\,d$ for $\vartheta = \pi/2$ (edge-on
impact).  The term $P(E,\vartheta)$ represents the total probability
for dissociation upon ion collision, for a particle with energy $E$
and incoming direction $\vartheta$. The ion collision rate is
calculated by convolution of the term $\Sigma(E)$ over the CR
spectrum $I_{\rm i}(E)$
\begin{eqnarray}\label{collRateIons_eq}
  R_{\rm i,\,CR}^{\rm coll} = 4\pi\int_{E_{\rm min}}^{E_{\rm max}}F_{\rm C}\,I_{\rm i}(E)\,\Sigma(E)\,{\rm d}E
\end{eqnarray}
Because each (electronic) ion interaction leads to the removal of 
two carbon atoms from the PAH, to obtain the C-atom ejection rate
the collision rate has to be multiplied by a factor of~2:
\begin{eqnarray}\label{ejectionRateIons_eq}
  R_{\rm i,\,CR} = 2 \times R_{\rm i,\,CR}^{\rm coll} = 
	           8\pi\int_{E_{\rm min}}^{E_{\rm max}}F_{\rm C}\,I_{\rm i}(E)\,\Sigma(E)\,{\rm d}E
\end{eqnarray}
The factor, $F_{\rm C}$, takes Coulombian effects into account (see below).
For $E_{\rm max}$ we adopt a value of 10 GeV, corresponding to the
highest energy for which experimental stopping determinations exist and
thus for which Eq. \ref{stoppingTotal_eq} is valid. Moreover the CR 
intensity and stopping power decrease rapidly, 
so ions with energy above 10 GeV do not contribute significantly to the integral in
Eq. \ref{collRateIons_eq}. Concerning the lower integration limit, we 
perform the calculation for $E_{\rm min}$ = 5 MeV/nucleon,
the lower limit assumed for the CR spectra.

For interactions with CR electrons, each binary collision 
results in the ejection of one single C-atom. Thus the ejection rate 
coincides with the collision rate, and are both given by the following
relation
\begin{eqnarray}\label{collRateElectrons_eq}
  R_{\rm e,\,CR} = R_{\rm e,\,CR}^{\rm coll} = 0.5\,N_{\rm C}\,
	           4\pi\int_{E_{\rm min}}^{E_{\rm max}}
	           F_{\rm C}\,I_{\rm e}(E)\,\sigma(E)\,{\rm d}E
\end{eqnarray}
where the ejection cross section \textit{per target atom} $\sigma$ has
to be multiplied by the number of C-atom in the
molecule, $N_{\rm C}$. 
The factor 0.5 takes the angle averaged orientation into
account (MJTa). As for the CR ions we assume as
integration limits the values $E_{\rm min}$ = 5 MeV and 
$E_{\rm max}$ 10 GeV. For the electrons, the upper limit is not 
constrained by the stopping theory (the expression for the ejection 
cross section holds for even higher energies) and is only related to 
the steepness of the spectrum which results in a negligible contribution 
from electrons with energy above 10 GeV (cf. Fig. \ref{CRspectra_fig}).

The Coulombian correction factor $F_{\rm C}$ (MJTb) takes into
account the fact that both target and projectiles are charged, and
that the collision cross section could be enhanced or diminished
depending on the charge of the PAH. Because we are considering high energy
ions and electrons, which are unaffected by the Coulombian field, 
$F_{\rm C}$ is always unity.

\section{Results}


\begin{figure*}
  \centering
 \includegraphics[width=1.\hsize]{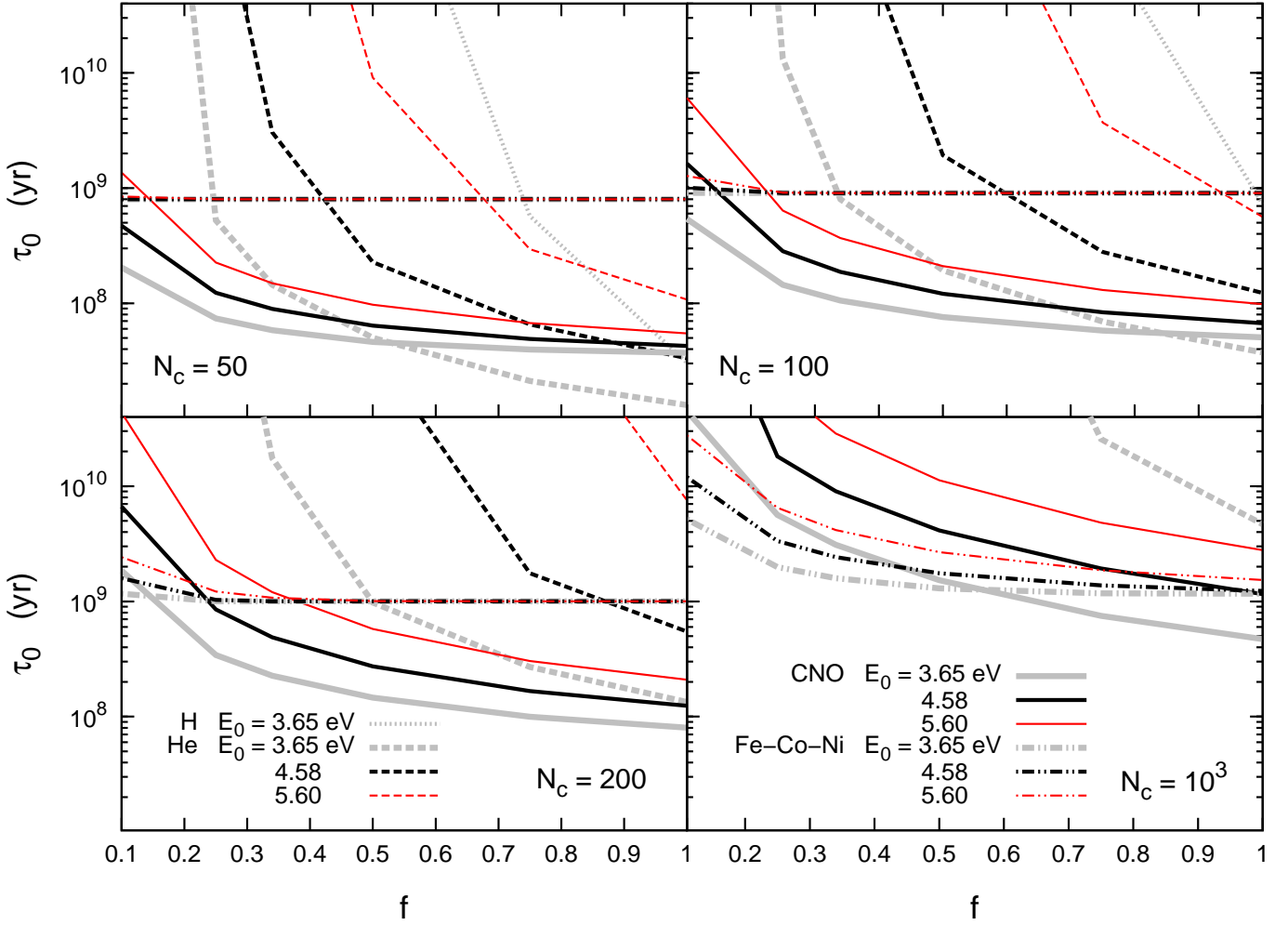}
  \caption{
	   PAH survival time against
           collisions with CR ions, as a function
           of the fraction $f$ of the transferred energy 
         that goes into vibrational excitation. The lifetime has 
           been calculated for PAHs of different sizes ($N_{\rm C}$), 
           assuming three values for the binding energy of the
         ejected fragment, $E_{\rm 0}$. The line style indicates the
         type of ion: dotted for H, dashed for He, solid for CNO and
         dot-dot-dashed for Fe-Co-Ni. The line color indicates the 
         value of the fragment binding energy: gray for $E_{\rm 0}$ =
         3.65 eV, black for $E_{\rm 0}$ = 4.58 eV and red for 
         $E_{\rm 0}$ = 5.6 eV. 
           }
  \label{ionTau0_fig}
\end{figure*}


\subsection{PAH lifetime}

Collisions with CR ions and electrons will cause a progressive decrease in the
number of carbon atoms in a PAH molecule. For interactions with ions, after 
a time $t$ this number is reduced to
\begin{eqnarray}\label{NtION_eq}
  N_{\rm C}(t)  = N_{\rm C}(0)\,-\,R_{\rm i,\,CR}\,t
\end{eqnarray}
and the fraction of carbon atoms ejected from this PAH is
\begin{eqnarray}\label{FlION_eq}
  F_{\rm L}(t) = \frac{R_{\rm i,\,CR}\,t}{N_{\rm C}(0)}
\end{eqnarray}
where $R_{\rm i,\,CR}$ is the C-atom ejection rate from 
Eq. \ref{ejectionRateIons_eq}. We assume that the PAH is destroyed
after the ejection of 1/3 of the carbon atoms initially present
in the molecule. This occurs after a time $\tau_0$ which we adopt
as the PAH lifetime against CR ions bombardment, and is given by
\begin{eqnarray}\label{timeConstant_ION_eq}
  \tau_{\rm 0}  = \frac{N_{\rm C}}{3\,R_{\rm i,\,CR}}
\end{eqnarray} 
For electron collisions, the number of carbon atoms in
the PAH molecule after a time $t$ is 
\begin{eqnarray}\label{NtELECTRON_eq}
  N_{\rm C}(t)  = N_{\rm C}(0) \exp \left[-t\,/\,\tau'\right]
\end{eqnarray}
and the fraction of carbon atoms ejected from this PAH is
\begin{eqnarray}\label{FlELECTRON_eq}
  F_{\rm L}(t) = \left(1 - \exp \left[ -t\,/\,\tau' \right] \right)
\end{eqnarray}
with the time constant $\tau'$ = $N_{\rm C}/R_{\rm e,\,CR}$,
where $R_{\rm e,\,CR}$ is the ejection rate from
Eq. \ref{collRateElectrons_eq}. The ejection of 1/3 of carbon 
atoms originally present in the PAH molecule, after which the
PAH is considered destroyed, takes the time $\tau_{\rm 0}$ given by
 \begin{eqnarray}\label{timeConstantELECTRON_eq}
  \tau_{\rm 0}  = \ln\left(\frac{3}{2}\right)\,\tau' = 
	          \ln\left(\frac{3}{2}\right)\,\frac{N_{\rm C}}{R_{\rm e,\,CR}}
\end{eqnarray}
As for ion collisions, we adopt $\tau_{\rm 0}$ as the PAH survival 
time against CR electrons. 
The ejection rate $R_{\rm e,\,CR}$
scales linearly with $N_{\rm C}$, hence the corresponding lifetime
$\tau_{\rm 0}$ is independent of PAH size.

For collisions with ions, we calculated the lifetime of PAHs of four
sizes $N_{\rm C}$ = 50, 100, 200 and 1000 (radius $a_{\rm PAH}$ =
6.36, 9, 12.7, 28.5 \AA, $n_{\rm max}$ = 10, 20, 40 and 200
respectively) bombarded by CR H, He, CNO and Fe-Co-Ni ions,
assuming three different values for the electronic binding energy,
$E_{\rm 0}$ = 3.65, 4.58 and 5.6 eV (cf. Sect. 4.1 in MJTb).
In principle, not all the energy transferred to the PAH
will be internally converted into vibrational modes, with consequent
relaxation through dissociation (or IR emission). Other processes can
occur, for instance the production of Auger electrons which will carry
away from the molecule a part of the transferred energy. We don't know
exactly how quantify this energy partitioning and so we introduce a
factor $f$, which represents the fraction of the transferred energy
$T$ that goes into vibrational excitation. 

The survival time as a function of the factor $f$ is shown
in Fig. \ref{ionTau0_fig}.
The PAH lifetime becomes shorter as more energy goes into vibrational
excitation (increasing $f$) and for lower values of $E_{\rm 0}$,
because this implies a higher temperature, $T_{\rm av}$, and a lower
energy to eject a fragment, 
resulting in a larger dissociation probability (Eq. \ref{totalProb2_eq}). 

The dissociation probability is more sensitive to $E_{\rm 0}$ when the
energy available for dissociation is lower, i.e. light projectile and
small $f$, and when the same amount of energy has to be spread over
more bonds (increasing size).  This explains why the separation
between the time constant curves corresponding to the different values
of $E_{\rm 0}$ decreases with increasing available energy and mass of
the projectile (from H to Fe-Co-Ni), while it gets bigger for larger
PAHs. Big PAHs are more resistent to CR bombardment because for any
given transferred energy their dissociation probability is lower.

The lifetime against Fe-Co-Ni bombardment is essentially constant -
except for very low available energy and very large molecules 
($N_{\rm C}$ = 1000).  Its large value, $\tau_{\rm 0} \sim$ few 10$^{9}$ yr,
results from the fact that, despite the huge amount of energy
transferred into the molecule (cf. Fig. \ref{ionStopping_fig}), the
Fe-Co-Ni abundance in CRs is small. This leads to a low
collision rate and long lifetime. For hydrogen, the high
abundance is not enough to compensate for the small stopping power,
which rapidly decreases above 1 MeV/nucleon. As a result the
lifetimes are long. Since the energy transferred to the
molecule is small, the collision rate and the survival times
are sensitive to the adopted parameters: the fraction of the 
energy transferred available for dissociation $f$, the
fragment binding energy $E_{\rm 0}$ and the PAH size. This
variability is shown in Fig. \ref{ionTau0_fig}. The situation for 
helium and CNO lies somewhere between the cases
for H and Fe-Co-Ni.

\begin{figure}
  \centering
  \includegraphics[width=1.\hsize]{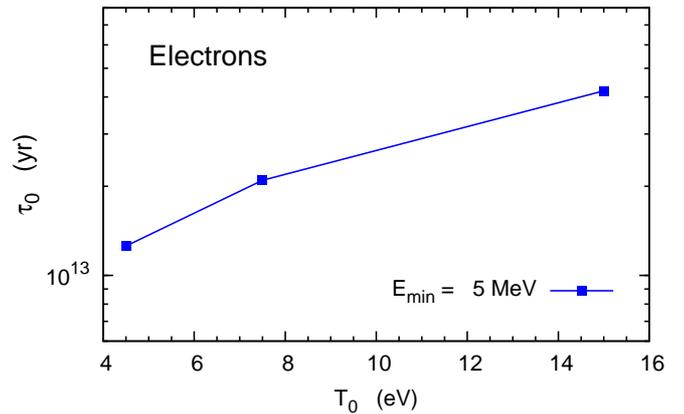}
  \caption{Time constant $\tau_{\rm 0}$ for carbon atoms removal due to 
           collisions with CR electrons, as a function of the 
           threshold energy $T_{\rm 0}$.
           }
  \label{tau0E_fig}
\end{figure}

The CR electron time constant is independent of the PAH size, and has
been calculated for three values of the threshold energy for carbon
atom ejection, $T_0$ = 4.5, 7.5 and 15 eV.
The results are 
shown in Fig. \ref{tau0E_fig} and summarized in Table \ref{tau0Elec_tab}. 
The PAH lifetime is longer for increasing values of $T_0$
because of the diminution in the
cross section $\sigma$.
Given that $\sigma$ is
almost constant for energies above $\sim$ 2 MeV, for a fixed
$T_0$ this part of the spectrum does not contribute to the variation.
Because of the small
ejection cross section and the steep CR electron spectrum, all the
calculated time constants are long ($\tau_0 >$10$^{13}$ yr).


\begin{table}
  \begin{minipage}[t]{\columnwidth}
    \caption
        {Time constant $\tau_{\rm 0}$ for carbon atom ejection following PAH
         collisions with CR electrons.
        }
        \label{tau0Elec_tab}      
        \centering          
        \renewcommand{\footnoterule}{}      
        \begin{tabular}{c c c c | c}     
          \noalign{\smallskip}
          \noalign{\smallskip}
          \hline\hline       
          \noalign{\smallskip}
                                  &  \multicolumn{3}{c}{$T_0$ (eV)}  &    \\
                                  &  4.5   &  7.5    &  15.          &  $E_{\rm min}$ (MeV)      \\  
          \noalign{\smallskip}
          \hline                    
          \noalign{\smallskip}         
           $\tau_{\rm 0}$ (yr)  &   1.2$\times$10$^{13}$     &  2.0$\times$10$^{13}$   &  4.2$\times$10$^{13}$   &  5   \\ 
          \noalign{\smallskip}
          \hline 
        \end{tabular}        
  \end{minipage}
\end{table}      


\subsection{Discussion of the uncertainties}

The first source of uncertainty affecting the study of PAH 
interaction with high energy particles is the accuracy of 
the stopping theory used to calculate the energy loss of 
ions in matter. This is a difficult issue because the current
theory, based on the Bethe-Bloch equation and described in Sect. 2.1,
is in fact a combination of different theoretical approaches 
with corrections coming from fits to the experimental data.
A better way to pose the problem is by instead asking: ``How accurately 
can stopping powers be calculated? '' The comparison between theory
and experiments, together with the evaluation of possible variation
sources such as structural variations in the targets, provides an accuracy
ranging from 5 to 10 \% \citep{ziegler99}.

In our calculation we do not use the complete Bethe-Bloch equation 
(implemented in the SRIM code with all corrective terms) but the 
analytical approximation given by Eq. \ref{stoppingApprox_eq}. The
discrepancy between the stopping power curves from these two 
formulations is very small, and introduces an uncertainty of at most 
10 \% in our calculation. An additional source of uncertainty 
comes from the calculation of the energy transferred to the PAH
using the approximation in Eq. \ref{eTransfIon_eq}, which assumes
a constant stopping power along the distance travelled through
the molecule. In this case the difference with the exact 
calculation is also limited to less than 10 \%.  

Concerning the collisions with high energy ions, the dominant source of
uncertainty in the determination of the PAH lifetime is the
fragment binding energy $E_{\rm 0}$ (cf. MJTb, Sect. 4.1), which
is  strongly modulated by other parameters: type
of ion, PAH size and fraction of energy available for dissociation,
as clearly shown in Fig. \ref{ionTau0_fig}. To give
an example, when $E_{\rm 0}$ goes from 3.65 to 5.6 eV,
the time constant for a 50 carbon atoms
PAH colliding with helium varies by a factor of $\sim$
2$\times$10$^{7}$ for $f$~=~0.25 and by a factor of 8 for $f$ = 1. For
CNO impacting on a 200 C-atom molecule the change is a factor of 7
and 3 for $f$~=~0.25 and $f$ = 1 respectively.
These numbers give an idea of the huge and complex variability
induced by the uncertainty in the parameter $E_{\rm 0}$.

In the treatment of PAH collisions with high energy electrons,
two sources of uncertainty have to be considered: the analytical
approximation to the numerical solution for the ejection cross section
(Eq. \ref{sigmaElectron_eq}) and the choice of the threshold energy $T_0$.
While the discrepancy
between the analytical expression and the numerical solution for $\sigma$ is
only a few percent \citep{mckin48}, the variation of $T_0$ from
4.5 to 15 eV implies a variation in the cross section of about a
factor of 9 around the peak (electron energy $\sim$ 0.1 MeV), and a
factor of 3 above $2$ MeV, where the cross section becomes almost
constant. This results in a change of the electron time constant of a
factor of $\sim$ 3.4,
consistent with the cross section variation. 
To summarize, in our study we explored a wide range in values for 
$T_0$ and $E_{\rm 0}$,
and the resulting variation is in some sense the maximum possible.

A final remark concerns the CR spectrum at low energies. As previously
mentioned, the major contribution to PAH destruction comes from
low-energy CRs, both because the energy transfer is more
efficient for low energy projectiles (ions) and the CR intensity
decreases with increasing energy (for both ions and electrons).
Unfortunately, because of the solar modulation this part of the
spectrum is not accessible to measurements from Earth and needs to be
evaluated theoretically. The result of this evaluation is not unique,
and various spectral shapes of the interstellar CR spectrum
have been shown to be consistent with the measured CR intensity
(cf. Sect. 4). The spectra we adopt, which of course match the
measured high energy spectra, have been calculated using a rigorous
physics and taking into account all possible constraints from
available measurements, including the stringent requirements from the 
CR ionization rate. Nevertheless the lack of data in such a
crucial CR energy region still represents
a source of uncertainty that is important to keep in mind, although it
does not have a significant impact at the level of accuracy for our study. 
For an overview of the current state of the art 
of the uncertainties in the CR fluxes we refer the reader to 
\citet{padovani09} and references therein.


\section{Discussion}

\subsection{The lifetime of PAHs against CR processing}

\begin{figure}
  \centering
  \includegraphics[width=1.\hsize]{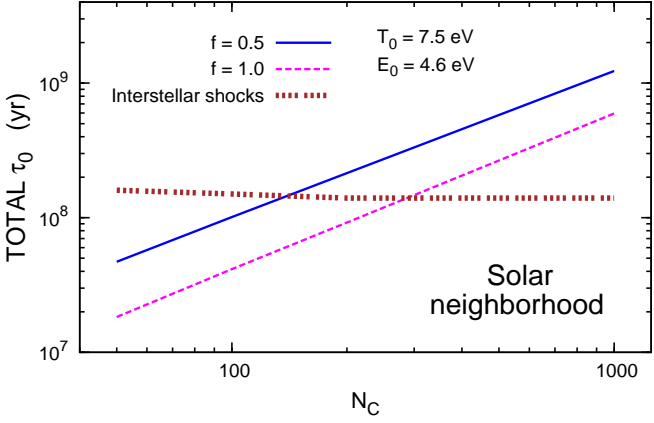}
  \caption{PAH survival time against CR bombardment (ions $+$ electrons)
           as a function of the molecule size ($N_{\rm C}$). The total
           lifetime has been calculated for $f$ = 0.5 and 1, where $f$ is the
           fraction of the transferred energy available for dissociation, and
           adopting our reference values for the threshold energy for carbon atom
           ejection, $T_0$ = 7.5 eV and for the fragment binding energy, $E_0$ =
           4.6 eV. We remind the reader of the variation in the calculated
           survival time against CRs, due to the uncertainty on the parameters
           $E_0$ and $T_0$ (cf. Fig. \ref{ionTau0_fig} and Fig. \ref{tau0E_fig}).
           The PAH lifetime against shock destruction in the ISM is shown for
           comparison.
           }
  \label{tau0ALL_fig}
\end{figure}

In Sect. 6.1, we estimated the PAH lifetime against CR bombardment
in the neighborhood of the Sun,
i.e. considering an interstellar spectrum based on galactic CR 
measurements near the Earth but corrected for the influence of the 
Heliosphere (cf. Sect. 4). 
Fig. \ref{tau0ALL_fig} shows the PAH lifetime against CR bombardment
(ions $+$ electrons) as a function of the size of the molecule,
$N_{\rm C}$, for $f$ = 0.5 and 1, assuming 
our reference values of 7.5 and 4.6 eV for the
parameters $T_0$ and $E_0$ respectively.
These lifetimes should be compared to the lifetime against shock
destruction of PAHs of approximately 150 Myr (MJTa). Depending
on the amount of transferred energy available for dissociation (50
or 100 \%), CRs are calculated to be the dominant
destruction agent in the warm ISM for PAHs with less then $\sim$150 
and $\sim$280 C-atoms respectively. For larger PAHs,
shocks take over. In this assessment, it should be kept in mind that
shock
processing occurs predominantly in warm intercloud medium of the ISM:
shocks faster than 100 km s$^{-1}$ are very rare in the cloud phase
since supernova remnants predominantly expand in the warm or hot
intercloud medium and the shock speed in clouds is then down by the
square root of the density ratio between these different phases in the
ISM \citep{jones94}. Shock processing of PAHs in diffuse clouds will
then predominantly occur when cloud material exchanges with the warm
intercloud medium (where shock timescales are short). The timescale
for this exchange is included in the lifetime estimate for shock
processing. CRs on the other hand penetrate all phases of the
ISM, except perhaps the densest molecular cloud cores, and can process
all material that they interact with. 

Finally, we note that the PAH survival times against CR electrons
($\tau_0 >$10$^{13}$ yr) are longer than the Hubble time, implying
that CR electron collisions are not important for the
processing of interstellar PAHs. We point out the fact that
this study does not include the secondary electrons with an energy of
about 30 eV produced by primary CR electrons by the initial ionization
of H or H$_2$ in diffuse clouds. Such electrons are energetic enough
to induce some PAH fragmentation but, although initially abundant, they
decay easily.


\subsection{PAHs in galactic halos}

ISO and Spitzer observations revealed that PAHs are abundant in
the halos of normal spiral galaxies \citep{irwin06, irwin07, whaley09}.  
Here, we will discuss the potential role of CRs in the 
destruction of these PAHs.

After having estimated the PAH lifetime against CR bombardment in the
vicinity of the Sun, the question that needs to be addressed is whether 
the CR spectrum changes across the galaxy and with the galactic 
latitude, and how this may affect the PAH lifetime. 

\subsubsection{The CR spectrum in galactic halos}

A powerful tool to probe the presence of CRs in regions
distant from the Solar System has been provided by gamma and radio
astronomy.  CR protons interact hadronically with the nuclei 
of the interstellar medium (mainly protons and $\alpha$ particles)
producing both charged and neutral pions \citep{fermi50, stecker73}. 
Charged pions decay into secondary leptons, while neutral pions decay 
into two $\gamma$ photons:
\begin{eqnarray}\label{pion_eq}
  p + N    &  \rightarrow  & X + \pi^{\pm/0} \\
 \pi^{\pm}  &  \rightarrow  & \mu^{\pm} + \nu_{\mu}/\bar{\nu}_{\mu}\; 
              \rightarrow e^{\pm} + \nu_{e}/\bar{\nu}_{e} + 
              \nu_{\mu} + \bar{\nu}_{\mu}  \\
   \pi^0   &  \rightarrow  &  2 \gamma
\end{eqnarray}
Electrons loose energy via synchrotron and inverse Compton, producing
detectable radio emission, while $\gamma$-rays are directly measurable
by gamma-ray telescopes. The quantity of cosmic-ray protons can then be 
calculated from their decay products, moreover, because photons are 
not affected by magnetic fields, they can provide a direct indication 
about the location of their sources \citep[e.g.][]{ramana93, hunter97}.

Different studies \citep[see][ and references therein]{shibata07} have
shown that the slope of the energy spectrum is almost independent of
the observational site in the Galaxy, but this is not true for the
amplitude of the spectrum. In the galactic plane the CR
intensity shows a longitudinal gradient \citep{shibata07}, with an
exponential decrease from the Galactic center toward the periphery
with a scale length of 20 kpc. Assuming the galactic parameters from
\citet{shibata07} we obtain an enhancement at the Galactic center of
about a factor 1.5 with respect to the Solar System.

Radio and gamma observations, together with the study of unstable CR
nuclei such as $^{10}$Be \citep{haya58, simpson88} have demonstrated that
CRs are not confined to the thin disk which contains their
sources, as assumed by the leaky-box model \citep{simpson88}, but are
able to travel long distances outside the galactic plane.  Various
models for CR propagation \citep[e.g.][]{ginzburg80, ginzburg88,
  bere90, strong98} assume a thin disk (half-thickness $h_{\rm d}
\sim$ 100 -- 200 pc) located on the galactic plane, where the sources
reside, surrounded by a large diffusive halo with half-thickness
$h_{\rm h} \sim$ 1 -- 15 kpc , where CRs spend part of their
life.

\begin{figure}
  \centering
  \includegraphics[width=1.\hsize]{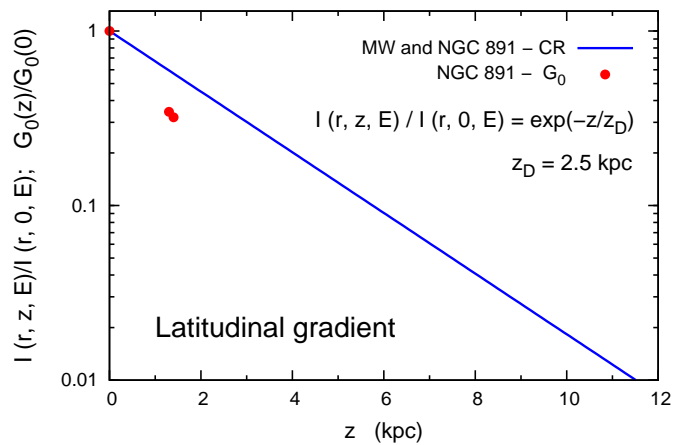}
  \caption{The latitudinal CR gradient in both Milky
           Way (MW) and NGC 891, as a function of the vertical distance $z$ 
           from the galactic plane together with the vertical
           variation of the interstellar radiation field $G_0$ in
           NGC 891 (\textit{courtesy F. Galliano}).
           The CR intensity 
           $I$ decreases with the scale height $z_{\rm D}$ of 
           the diffusion coefficient.
           }
  \label{latGrad_fig}
\end{figure}

Using the formalism developed by \citet{shibata07III} the latitudinal
gradient of the CR intensity for a given ion $i$ can be written as
\begin{eqnarray}\label{latGrad_eq}
  \frac{I_{i}(r, z, E)}{I_{i}(r, 0, E)} \simeq \exp\left(-|z|/z_{\rm D}\right)
\end{eqnarray}
where $r$ is the radial distance from the center of the galaxy projected 
on the galactic plane
and $z$ is the latitudinal distance from the Galactic plane.
Eq. \ref{latGrad_eq} shows that the CR intensity attenuates with 
the scale height $z_{\rm D}$ of the diffusion coefficient. The latitudinal
gradient is calculated assuming the value $z_{\rm  D}$ = 2.5~kpc proposed by 
e.g. \citet{shibata07} and is shown in Fig. \ref{latGrad_fig}.

Ultimately, CRs are thought to be accelerated by supernova
remnants, tapping some 10\% of the SN energy. For other galaxies, we
will scale the overall CR density with the star formation rate
of the galaxy and we will adapt scale lengths and scale heights
appropriate for the \object{Milky Way}. Since the CR intensity decreases
exponentially outside the galactic plane, we expect the PAH lifetime
at higher latitudes to be enhanced by the same factor.


\subsubsection{The origin and evolution of PAHs in galactic halos}

PAH emission has been observed at high latitudes in several normal
galaxies with scale heights of 2-3 kpc \citep{irwin06, irwin07, whaley09}. 
The study of NGC 891, the edge-on twin of the Milky Way, is 
particularly instructive. While the vertical distribution of the 8 $\mu$m
PAHs emission is noticeably narrower than that of the cold dust
radiating at 450 $\mu$m, modeling reveals that this likely reflects the
variation of the stellar population from the disk to the halo; e.g.,
the dominant (FUV) heating sources of the PAHs are located in the disk
while the old stellar population, which contributes substantially to
the heating of the dust, extends some 3 kpc above the disk \citep{whaley09}.

Some of the PAH molecules, as well as the dust, present at high
galactic latitudes may originate from mass losing AGB stars in the
halo. For the Milky Way, recent optical and infrared surveys (2MASS
and DENIS) have revealed the presence of C-rich AGB stars enshrouded
in dusty ejecta at distances of 2-6 kpc above the plane \citep{gro97,
 mauron08}. Some 100 such C-stars are known in the halo – up to
distances of 150 kpc – and many of these belong to the tidal stream of
the Sgr dwarf galaxy \citep{mauron05}. There is no complete census of such
objects and a mass balance of PAHs and dust in the halo cannot yet be
assessed. 

More likely, though, most of the PAHs and dust
represent signpost of the large scale circulation of matter between
the disk and halo. Indeed, the vertical distribution of PAHs is
similar to that of the dust –- once the differences in heating are
accounted for –- and to other tracers of this circulation pattern such
as the diffuse ionized gas. If PAHs are indeed transported to high
latitudes through the action of the galactic fountain, then the
acceleration associated with this process must have been very gentle
since PAHs are quickly destroyed in shocks faster than 100 km s $^{-1}$
(MJTa). Likely, PAHs are entrained as cloud(let)s
sheared off the chimney walls in the hot gas of the
venting supernovae and lifted high above the plane. As shown
in MJTb, PAHs are rapidly destroyed in hot gas. Hence, after evaporation of
these cloudlets driven by thermal conduction from the hot gas, PAHs
will be gone and clouds subsequently condensing through thermal
instabilities in the halo will not show up in the PAH emission bands.

Hence, PAHs in the halo may represent a `molecular dye' with
which this entrainment aspect of the disk-halo circulation flow can be
separated from the evaporation/condensation pattern and followed in
detail. The effectiveness of this PAH-dye is modified on the one hand
by the diffusion of the FUV photons needed for their excitation from
the disk and on the other hand by destruction of PAHs in the harsh
environment of the halo. Here, we are concerned with the latter
aspect, the lifetime of PAHs at high latitudes above the plane. 
First, the residence time is of interest. 
With a total mass of Warm Intercloud material of $\sim$10$^9$ M$_{\sun}$ 
in the lower ($\sim$ 0.5 kpc) halo and a circulation rate of 5
M$_{\sun}$/yr between the plane and the halo for the Milky Way, the
residence time of PAHs at these latitudes is some 2$\times$10$^8$
yrs. The residence time at the higher latitudes ($\sim$ 2 kpc) we are
interested in here will be commensurately larger.
At the latitudes where PAHs are observed ($\sim$2 kpc), we expect 
that supernova shock waves, an important
destruction agent in the disk of the galaxy, are of little
concern.
UV photolysis is generally considered to be a main agent for the
destruction of small ($<$50 C-atoms) PAHs, weeding out the less stable
(e.g., smallest and/or non-compact) PAHs on a rapid timescale
\citep{tielens08}. Indeed, the minimum size in the
PAH-size-distribution is thought to reflect this process.  However,
because this process is very sensitive to size, compact PAHs only
slightly larger than this minimum size are essential
`indestructible'. Thus, we expect that PAHs transported upwards from
the plane by the galactic fountain have already been weeded down to
the most stable forms and because of the decreased UV flux with
latitude (Fig. \ref{latGrad_fig}), further UV photolysis will be of
little concern.
Using the results of this paper, we can estimate the destruction of
PAHs by CRs. 

Taking the CR distribution in the halo
from Sect. 7.2.1 
we calculate the CR destruction timescale. 
Because of its similarity with the Milky Way, we adopt for NGC 891 
the same diffusion scale height $z_{\rm D}$ = 2.5 kpc.
This implies that the latitudinal gradient from Eq. \ref{latGrad_eq}
is the same for both galaxies.
The CR intensity variation in NGC 891 with respect to the solar
neighborhood is given by the following expression
\begin{eqnarray}\label{crNGC891_eq}
  \frac{I_{i, \rm NGC 891}(0, |z|, E)}{I_{i, \rm MW}(r_{\sun}, 0, E)} = 
       (1.5\times\frac{3.8}{3})\,\exp\left(-|z|/z_{\rm D}\right) 
\end{eqnarray}
where 1.5 is the CR enhancement factor in the galactic center with
respect to the solar neighborhood and the factor 3.8/3 takes into
account the increased cosmic-ray intensity in NGC 891 due to its
higher star formation rate \citep[3.8 M$_{\sun}$ yr$^{-1}$,][]{popescu04} 
with respect to the Milky Way \citep[$\sim$3 M$_{\sun}$ yr$^{-1}$,][]
{scalo86, pran98}. At the typical PAH detection scale height 
$|z|$ =  3 kpc we obtain a decrease of the CR intensity of a factor 
of 0.6. The lifetime against destruction will be then enhanced by a
factor of (1/0.6). 

\begin{figure}
  \centering
  \includegraphics[width=1.\hsize]{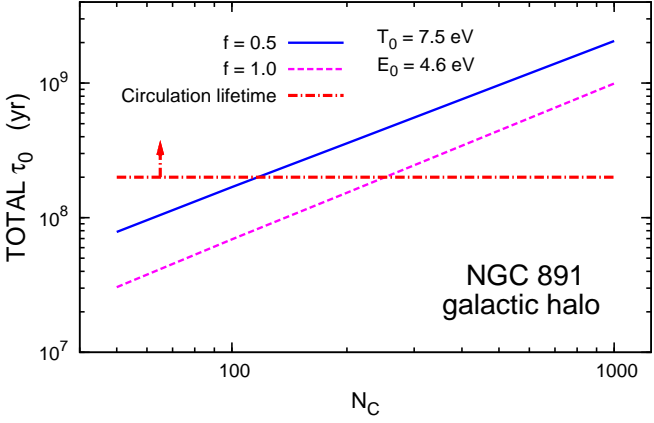}
  \caption{Total PAH lifetime against CR bombardment in the
           halo of NGC 891, evaluated at an altitude of 3 kpc
           from the galactic plane, compared with the circulation
           timescale between the disk and the halo. The value of
           the circulation lifetime represents a lower limit, as
           indicated by the arrow. Otherwise as Fig. \ref{tau0ALL_fig}.
           }
  \label{tau0ALLhalo_fig}
\end{figure}

The total PAH lifetime in the halo of NGC 891 compared with the
circulation timescale is shown in Fig. \ref{tau0ALLhalo_fig}.
Depending on the amount of energy available for dissociation,
the survival time against CRs for PAHs with more than $\sim$100~C~-~atoms
is comparable or longer than the 
circulation lifetime.
PAHs could be good tracers of material entrained in the
plane and carried to high latitudes by the galactic fountain as long
as this material never loses its `identity' through evaporation into
the hot gas followed by recondensation into cloudlets. Of course, as
emphasized by \citet{whaley09}, such a study has to properly evaluate
the flux of the pumping of UV photons in order to be of quantitative
value. In that case, mid-IR studies can potentially trace this
entrainment process and the exchange between cool and hot phases on
arcsecond size scales. Nevertheless other factors have to be taken into
account, such as the possibility of local PAH production in the halo.


\subsection{PAHs in galactic winds}

Starburst galaxies are characterized by very intense star formation,
usually concentrated in the nuclear region of the galaxy, which can
occur as a single instantaneous burst or as a high star formation 
activity sustained over a period of time.
Since galactic supernovae
are considered the most probable acceleration source for CRs
with energy below the ``knee'' \citep[few 10$^{15}$ eV - see e.g.][]
{drury94}, we expect that an enhancement of the SFR will result in a
corresponding increase of the CR intensity and dust destruction
by shocks.

In our previous work (MJTb) we discussed the observation of PAH
molecules in the bipolar outflow of the starburst galaxy M82
\citep{armus07, galliano08}. PAHs have been detected far
outside the galactic plane at least up to 6 kpc from the disk 
\citep{engel06}, in a region which is spatially correlated
with an extended X-ray emission originating from tenuous hot gas
(electron density $n_{\rm e}$ = 0.013 cm$^{-3}$, temperature $T$ =
5.8$\times$10$^6$ K). Our study has demonstrated that, under these
conditions, PAHs are destroyed by collisions with thermal electrons on
a timescale of few thousand years.  Our conclusion was that the PAH
survival is possible only if the molecules are isolated from this hot
gas, probably in cooler cloudlets of PDR-type gas entrained in the galactic wind.

In fact supernova-driven galactic-scale winds, often called
``superwinds'', are commonly observed in starburst galaxies, including
M82 \citep[see e.g.][ and references therein]{strick09}. In addition,
theoretical studies \citep{ipavich75, bre91, bre93, zira96, ptu97}, 
together with radio \citep{heesen09} and X-ray observations \citep{everett08} 
suggest the possibility of a CR-driven galactic wind, i.e. the 
possibility of ``bulk'' transport of CRs outwards into the halo, 
in addition to the diffusive transport predicted by CR propagation 
models (cf. Sect. 7.1).

Assuming also that the M82 superwind could be partially CR-driven, we  
can try to evaluate the CR intensity at the high latitudes where PAHs
are detected, to assess PAH lifetimes against CR bombardment
and compare them with the survival time in hot gas of the outflow.
Unfortunately the properties of superwinds at high galactic latitudes
are not well constrained. Theoretical studies mainly concern small
regions around the nucleus of the galaxy (radii of few hundred pc
from the center), where the transport of matter is dominated
by advection \citep{volk03, strick09}, while in the outer regions diffusion
is expected to be dominant \citep[e.g.][ for the starburst galaxy 
\object{NGC 253}]{zira06}. This is not the case in the Milky Way, where the
CR transport is diffusive on a kpc scale and becomes convective 
only at larger distances \citep{ptu97}. 

On the other hand, the CR propagation in M82 is
not yet totally understood. 
Star formation, and thus SNRs which re-presents the main CR sources, 
are concentrated in the small nuclear region from which the galactic 
superwind originates, we could then expect CRs being advectically
transported by the wind out of the galactic plane up to a certain height 
($\sim$ few hundred pc), and diffusing after in the volume occupied 
by the superwind, which is roughly a cylinder flaring at high $z$ with 
height of $\sim$ 12 kpc on the North side and 7.5 kpc on the South side 
\citep{lehnert99, stevens03}. 
It would also be possible that CRs occupy a spheroidal 
halo as in the Milky Way, maybe diffusing back to the disk after 
leaving the wind region. 

Because of this lack of information, especially on the confinement
volume for the CRs, what we present here is a 
toy-model based on some simplifying assumptions, but which 
could be useful to interpret PAH observations in the outflow
of M82. To calculate the CR spectrum in M82 we adopt
the same spectral index as in the Milky Way ($\gamma$ = 2.7)
and we assume that in the starburst nucleus the CR intensity scales
with the supernova rate $\nu_{\rm SN}$  (SN yr$^{-1}$) which is directly 
related to the star formation rate in the galaxy.
Taking the values $\nu_{\rm SN}(\rm M82)$ = 0.11 SN yr$^{-1}$ \citep{huang94}
and $\nu_{\rm SN}(\rm MW)$ = 0.02 SN yr$^{-1}$ \citep{diehl06}
we obtain $I_{\rm M82}(E)/I_{\rm MW}(E)$ = 
$\nu_{\rm SN}(\rm M82)$/$\nu_{\rm SN}(\rm MW)$ = 0.11/0.02 = 5.5.

Since PAHs are detected at vertical distances which are large compared
with the typical height of the convective zone, we neglect this latter
and assume pure diffusive CR transport above the galactic plane. We
adopt the exponentially-decreasing CR latitudinal gradient from 
Eq. \ref{latGrad_eq} \citep{shibata07} with $z_{\rm D}$ = $h_{\rm h}$/3,
where $h_{\rm h}$ is the maximum distance that CRs can reach 
above the plane through diffusion. Since the exact confinement region
in M82 is not known, we just take $h_{\rm h}$ equal to the vertical
height of the galactic superwind (12 kpc in the North side, 7.5 in the
South side), i.e. we assume that CRs can occupy at least as the same volume
than the gas from the outflow. 

The PAH lifetime against CR bombardment at the vertical distance $|z|$
above the starburst nucleus of M82 is then given by the following
expression
\begin{eqnarray}\label{tau0_M82_eq}
  \tau_{\rm 0, M82}(0, |z|) = \frac{1}{5.5\times1.5}\,\frac{\tau_{\rm 0, MW}(r_{\sun}, 0)}
                            {\exp\left(-|z|/z_{\rm D}\right)}
\end{eqnarray}
where 5.5 and 1.5 are the two enhancement factor of the CR intensity in
M82 with respect to the Milky Way (5.5) and in the galactic
center with respect to the solar neighborhood (1.5), which will imply a
decrease of the same factor for the PAH lifetime with respect to the 
survival time $\tau_{\rm 0, MW}(r_{\sun}, 0)$ in the vicinity of the Solar
System (Fig. \ref{tau0ALL_fig}). On the other hand 
the latitudinal gradient will result in a corresponding increase of 
$\tau_{\rm 0, M82}$ for increasing $|z|$, with scale height 
$z_{\rm D}$(North) = 4 kpc and $z_{\rm D}$(South) = 2.5 kpc.


\begin{figure}
  \centering
  \includegraphics[width=1.\hsize]{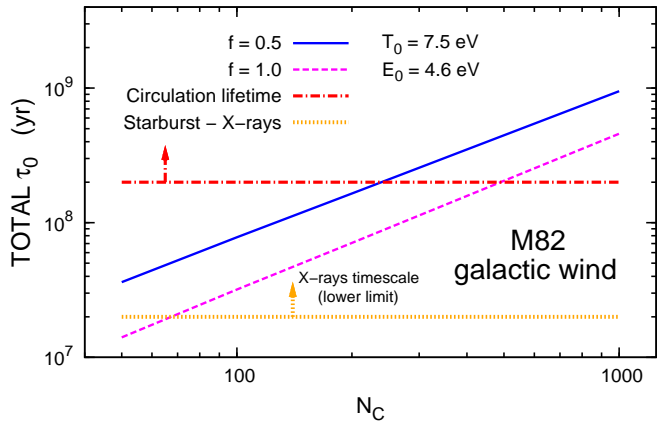}
  \caption{Total PAH lifetime against CR bombardment in the
           outflow of the starburst galaxy M82, evaluated at an 
           altitude of 6 kpc from the galactic plane. The PAH 
           survival time is compared with the circulation,
           starburst and X-rays timescales. The values of the
           circulation and X-rays timescales represent lower limits,
           as indicated by the arrows. Otherwise as
           Fig. \ref{tau0ALL_fig}.
           }
  \label{tau0ALLM82_fig}
\end{figure}


Let us calculate the PAH lifetime at the maximum distance $|z|$ = 6 kpc
where the molecules have been detected. This is done by scaling the time
constants $\tau_{\rm 0, MW}(r_{\sun}, 0)$ in Fig. \ref{tau0ALL_fig} by the
factor $A_{\rm }\sim$1.3 (averaged over the two hemispheres) and the 
result is shown in Fig. \ref{tau0ALLM82_fig}.
The PAH lifetime against CRs is longer than the starburst
  timescale (tens of Myr), except for the smallest molecules when 
  $f$ = 1, and for large PAHs ($N_{\rm C} >$ 150-200)
it is comparable or longer than the circulation lifetime ($\sim$ 200
  Myr).
In MJTb we showed that in the outflow of M82 PAH destruction due to
X-ray absorption would take place on a timescale of at least $\sim$ 20 Myr
(assuming that each X-ray photon absorption will lead to fragmentation).
This is comparable to or larger than the starburst lifetime.
PAHs dispersed in the hot gas will be rapidly destroyed by
collisions with electrons. To survive they need to be isolated from
this hot gas, presumably in a cooler and denser gas entrained in the
superwind. On the other hand CRs and X-rays can penetrate in those clumps,
thus the PAH lifetime there will be determined by the cosmic 
radiation and/or X-rays, depending on the PAH size and on the effective
value of the survival time against X-rays (cf. \S~6.1 in MJTb).

\subsection{PAHs in galaxy clusters}

We consider now the PAH processing by CRs in an extragalactic
environment, focusing on the specific case of cooling flow galaxy
clusters. The measurements of high-energy gamma and neutrino radiation
produced by the interaction of CRs (mainly protons) with the
intra-cluster medium (ICM) have demonstrated that galaxy cluster are
able to confine CRs for cosmological times \citep{bere97}. Our choice
is thus motivated by the reasonable expectation to find a
non-negligible population of CRs in such environments, although the
spectral index and amplitude are not well constrained by observations.

Many clusters of galaxies with cooling flows host a central galaxy
which often shows nuclear activity. Many papers support the presence
of CR protons in the relativistic plasma bubbles produced by
Active Galactic Nuclei (AGN), where the energetics are sufficient
to inject a significant amount of CRs into the ICM
\citep[see][ and references therein] {pfrommer04}. The
slope of the injected spectrum depends on the fraction of particles
released from the plasma. If all the CR content is ejected by the
radio plasma, we expect to find a moderately flat spectrum
($\alpha_{\rm inj} \approx$ 2.5) consistent with the flat spectra of
CR electrons indicated by the radio emission from radio
galaxies. On the other hand, if only a small
fraction of the CR population is able to leave the plasma via
diffusion, we expect to find an even flatter spectrum ($\alpha_{\rm inj} 
\approx$ 2.2) because the escape probability increases with momentum 
\citep{ensslin03}.

Another possible source of CRs in galaxy clusters is
represented by the galactic winds from the central galaxy transporting
CRs accelerated within the galaxy itself. Because Supernova Remnants
(SNRs) are believed to be the main sources of Galactic CRs
\citep[e.g.][]{ginzburg93, drury94}, the expected injection spectral index is
$\alpha_{\rm inj} \approx$ 2.4, if no further re-acceleration occurs.
Apart from protons, the principal constituents of Galactic CRs 
are  helium, carbon and iron ions. 

The CR population in galaxy clusters will then result from the
combination of the above-mentioned sources, with a spectral index
$\alpha_{\rm inj}$ which reflects its composite origin. After
injection from the source, CRs start diffusing away from the central
AGN. For the treatment of the transport of CR \textit{protons}
through the thermal ICM, we refer to the model elaborated by \citet
{pfrommer04}. The CR proton distribution function is described by
a power-law in energy (or momentum) where the injection spectral 
index is modified by the diffusion through the intra-cluster gas 
\begin{eqnarray}\label{pDensity_eq}
  g(r, E) = \frac{\tilde{n}_{\rm CR,\,0}}{\rm {GeV}}\,
  \left(\frac{r}{h_{70}^{-1}\,\rm{kpc}}\right)^{-1}\,
  \left(\frac{E}{\rm {GeV}}\right)^{- \alpha_{\rm p}}
\end{eqnarray}
with $\alpha_{\rm p} = \alpha_{\rm inj}\,+\,\alpha_{\rm diff}$. The
coefficient $\alpha_{\rm diff}$ results from the fact that diffusion
is momentum-dependent, and equals $\approx$ 1/3 for active diffusive
transport of CR protons in a Kolmogorov-like spectrum of hydro-magnetic
turbulence \citep{kol41}.
The model assumes the
standard $\Lambda$CDM cosmology with $H_0$ = 70 $h_{70}$ km s$^{-1}$
Mpc$^{-1}$, where $h_{70}$ indicates the scaling with $H_0$. The
normalization factor $\tilde{n}_{\rm CR,\,0}$ is the crucial parameter
required to determine the CR spectrum.

\citet{pfrommer04} constrained $\tilde{n}_{\rm CR,\,0}$ for a sample
of nearby cooling flow galaxy clusters with the aid of $\gamma$-ray
observations, using the fact that CR protons interact
hadronically with the thermal intra-cluster gas producing both charged
and neutral pions (see Sect. 7.2.1). Charged pions decay into secondary
electrons/positrons + neutrinos/antineutrinos, while neutral pions
decay into two $\gamma$'s ($\pi^0 \rightarrow 2 \gamma$). The amount
of cosmic protons can then be calculated from their decay products,
which are detectable in the $\gamma$-ray and radio bands.

It is important to remember here that the threshold energy for pion 
production is $E_{\rm th}$ = 0.78 GeV, implying that only the proton 
population with energy exceeding the threshold is able to produce
pions hadronically, and can then be constrained by $\gamma$-ray and 
radio observations. At lower energies the dominant energy loss mechanism
for CR protons is Coulombian diffusion on the electrons in the 
plasma (electronic excitation), which results in the depopulation of the
CR energy distribution with consequent modification of the
power-law spectrum \citep{mann94}.

Unfortunately, it is not possible to observationally constrain the
low-energy CR population in galaxy clusters, thus following
the argument from \citet{nath94} we decided to extrapolate
Eq. \ref{pDensity_eq} down to a minimum energy $E_{\rm min}$
representing the characteristic energy scale at which the spectrum
deviates from a power-low. A similar approach is used by
\citet{pfrommer04} as well. For the cut-off energy we adopt the value
$E_{\rm min}$ = 50 MeV, consistent with the range $\sim$ 50 - 75 MeV
found by \citet{nath06} in their study of the production of $^6$Li
by CR protons in galaxy clusters.  $E_{\rm min}$ represents
the lower energy that an energetic proton has to reach before
suffering large Coulombian losses, and can be equivalently expressed
in terms of the quantity of matter to be traversed in order to
experience the same Coulombian losses.

As previously mentioned, we assume the presence in the cluster CR 
population of the main components of the Galactic CRs other than
hydrogen: helium, carbon and iron.
The distribution function (\ref{pDensity_eq}) for the heavier elements 
is obtained by simply scaling the proton distribution function $g(r, E)$ with
the abundance with respect to the hydrogen $\chi_{i}$ of the corresponding 
ion $i$ = H, He, C, Fe
\begin{eqnarray}\label{ionDensity_eq}
  g_{i}(r, E) = \chi_{i}\,g\left(r, E\right)
\end{eqnarray}
where $E$ is the \textit{total} kinetic energy of the CR ion.
For the CR abundances 
we assume the standard values of the galactic CRs: 
$\chi_{\rm H}:\chi_{\rm He}:\chi_{\rm C}:\chi_{\rm Fe}$
= 1 : 0.1 : 10$^{-2}$ : 10$^{-3}$ \citep{maurin03}. The low-energy
cut-off for the heavier ions has been derived in the same way as for
Galactic CRs, i.e. scaling the proton value 
$E_{\rm min}^{\rm p}$ = 50 MeV by the ion atomic mass, 
$E_{\rm min}^i$ = $M_{1, i}\,E_{\rm min}^{\rm p}$, but different
parametrizations are also possible \citep{nath94}.

The CR intensity $I(E)$ (to which we usually refer as
``spectrum'') is related to the distribution function by the
following expression
\begin{eqnarray}\label{clusterSpectrum_eq}
  I_{i}(E) = \frac{v}{4\pi}\,g_{i}(r_{\rm c}, E)
\end{eqnarray}
where $v = \beta c$ is the velocity of the ion. The distribution
function $g_{i}$ is calculated for a given distance $r_{\rm c}$
from the center of the cluster. We adopt the value corresponding
to the cooling radius of the cluster, which defines the region
where the majority of the hadronically-generated $\gamma$-ray
luminosity originate.   

When the CR intensity is known, we can estimate the total lifetime against
CR bombardment, $\tau_{\rm 0}$. 
As an example, we calculate $\tau_{\rm 0}$
in two nearby cooling flow clusters from the \citet{pfrommer04}
sample, \object{A85} and \object{Virgo}, whose main properties are listed in Table
\ref{cluster_tab}.  A85 is one of the farther cluster in the sample,
with low electron density $n_{\rm e}$, large core radius $r_{\rm c}$
and high central temperature $T_0$.  Virgo is the closest cluster to
us, with quite high electron density, very small core radius and high
central temperature.

\citet{pfrommer04} calculated $\tilde{n}_{\rm CR,\,0}$ for different values
of the proton spectral index $\alpha_{\rm p}$. We adopt the value corresponding
to $\alpha_{\rm p}$ = $\alpha_{\rm inj}$ + $\alpha_{\rm diff}$ = 2.7, with
$\alpha_{\rm diff} \approx$ 1/3, because is consistent with the spectrum 
of the CR sources (AGN + SNRs) and allows a direct comparison with Galactic
CRs.


\begin{table}
  \begin{minipage}[t]{\columnwidth}
    \caption
        {Relevant parameters and CR normalization factor 
         $\tilde{n}_{\rm CR,\,0}$ for the two cooling flow
         clusters A85 and Virgo from the \citet{pfrommer04}
         sample.
        }
        \label{cluster_tab}      
        \centering          
        \renewcommand{\footnoterule}{}      
        \begin{tabular}{c c c c c c}     
          \noalign{\smallskip}
          \noalign{\smallskip}
          \hline\hline       
          \noalign{\smallskip}
                   &       &  $r_{\rm c}$             &  $n_{\rm e}$                   &
           $T$     &  $\tilde{n}_{\rm CR,\,0}^{a}$  \\
          Cluster  &  $z$  &  ($h_{70}^{-1}\,\rm kpc$)  &  ($h_{70}^{1/2}\,\rm cm^{-3}$)    &  
           (10$^{7}$ K)      &  ($h_{70}^{1/2}\,\rm cm^{-3}$)  \\  
          \noalign{\smallskip}
          \hline                    
          \noalign{\smallskip}         
          A85$^{b}$    &  0.0551  &  45   &  3.08$\times$10$^{-2}$  &  6.4  &  9.9$\times$10$^{-5}$    \\
          Virgo$^{c}$  &  0.0036  &  1.6  &  1.5$\times$10$^{-1}$   &  1.4  &  4.2$\times$10$^{-7}$     \\
          \noalign{\smallskip}
          \hline 
          \noalign{\smallskip}
        \end{tabular}        
  \end{minipage}
  ($a$): From \citet{pfrommer04}, with $\alpha_{\rm p}$ = 2.7. \\
  ($b$): Parameters from \citet{mohr99} and \citet{oe01} \\
  ($c$): Parameters from \citet{mat02} and \citet{ebeling98}. \\ 
\end{table}      

\begin{figure}
  \centering
  \includegraphics[width=1.\hsize]{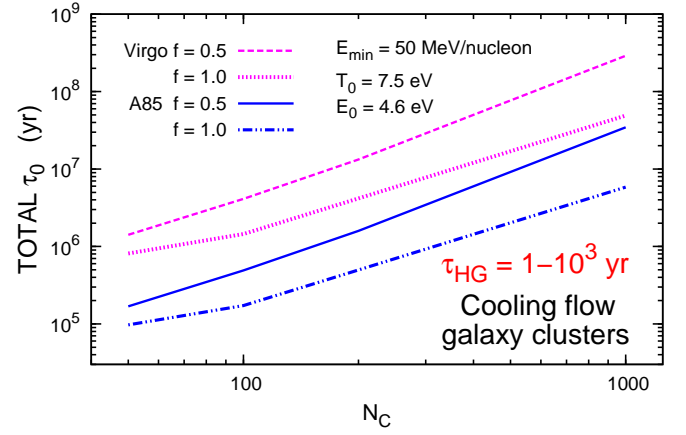}
  \caption{Total PAH lifetime against CR bombardment in the
           cooling flow galaxy clusters A85 and Virgo, compared with
           the survival time, $\tau_{\rm HG}$, in a hot gas. The value adopted
           for the lowest energy of the CR spectrum, E$_{\rm
             min}$, is 50 MeV/nucleon (see text), otherwise as
           Fig. \ref{tau0ALL_fig}.
         }
   \label{tau0ALLcluster_fig}
 \end{figure}

The results of our calculation are shown in Fig.\ref{tau0ALLcluster_fig}. 
In the galactic clusters under examination the cosmic-ray density is
enhanced with respect to the solar neighborhood, resulting in lifetimes
which are shorter by about 2-3 order of magnitude (cf. Fig. \ref{tau0ALL_fig}).
The longer time constants obtained for Virgo are the consequence of
its lower CR density with respect to A85.


In our previous work (MJTb) we showed that collisions with thermal
electrons and protons represent an efficient destruction mechanism 
for PAHs immersed in an ionized gas of $\sim$ 10$^6$ K.
Since the ICM is a moderately tenuous hot plasma, it would
be not surprising to find a quite short lifetime $\tau_{\rm HG}$ for 
the PAHs dispersed in the medium, due to such collisions.
We calculate the PAH lifetime in the hot intra-cluster medium of A85
and Virgo using Eq. 27 in MJTb and the temperatures and densities
listed in Table \ref{cluster_tab}.
In both clusters, PAHs embedded in the hot gas are rapidly destroyed on 
timescales of 1 -- 10$^3$ yr. In cold embedded gas, the lifetime of PAHs is 
set by CRs to be $\sim$10$^6$ -- 3$\times$10$^8$ yr (Virgo) and
$\sim$10$^5$ -- 3$\times$10$^7$ yr (A85).

The existence of PAHs in the intra-cluster medium is not
yet well constrained observationally. In fact to the best of
our knowledge a detection has not been reported, but hopefully
future observations will be able to provide more insights. 
Many of the quantities involved in our calculation are not well 
constrained, in particular the exact amplitude, shape and composition of the
CRs spectrum in the cluster and the value of the parameters
$E_0$ and $T_0$. 
Because the PAH lifetime in galaxy clusters set by CRs
is still short, a replenishing mechanism would be
required to explain their observation in such environments. We face
here another example of the well-known conundrum about the discrepancy
between the PAH/dust lifetime and the injection time scale into the
interstellar medium (2.5$\times$10$^9$ yr).

\section{Conclusions}

We have extensively investigated the stability of PAHs against
CR ions (H, He, CNO and Fe-Co-Ni) and electron bombardment in
both galactic and extragalactic environments. We consider CR 
particles with energy between 5 MeV/nucleon and 10 GeV. Collisions can lead
to carbon atom ejection, with a consequent disruption and destruction
of the molecule. The effects of CRs were then compared
with the other destruction mechanisms discussed in our previous
works: PAH processing by shocks with velocities between 50 and
200 km s$^{-1}$ (MJTa) and collisions with thermal ions and electrons
in a hot gas ($T$ = 10$^3$ -- 10$^8$ K, MJTb).

An ionic collision consists of two simultaneous processes which can be
treated separately: a binary collision between the projectile ion and
a single atom in the target (nuclear interaction) and energy loss to
the electron cloud of the molecule (electronic interaction). In the
high-energy regime considered here the nuclear stopping is totally 
negligible, and the energy loss process is dominated by the electronic
interaction, well described by the Bethe-Bloch equation.
The interaction of PAHs with high-energy electrons can be treated
in term of a binary collision between the incident electron and
a single nucleus in the target. 

The CR spectra we adopt in the solar neighborhood are 
based on measurement near the Earth but corrected for the
influence of the Heliosphere (solar modulation). To estimate
the CR variation across the disk and in the galactic halo we
adopt specific models based on $\gamma$-ray measurements.
In external galaxies we scale the overall CR density with 
the star formation rate of the galaxy, adapting scale lengths
and scale heights appropriate for the Milky Way.

We find that the timescale for PAH destruction by CR ions
depends on the electronic excitation energy $E_0$
and the amount of energy available for
dissociation. Small PAHs are destroyed faster, with He and the CNO
group being the more effective projectiles. 
CRs are able to process PAHs in diffuse clouds, where the destruction
due to interstellar shocks is less efficient. For electron
collisions, the lifetime is independent of the PAH size and varies
with the threshold energy $T_0$. The minimum lifetime 
is 1.2$\times$10$^{13}$ yr, longer than the Hubble time. 
Such a long timescale excludes
CR electrons as an important agent for PAH destruction. 

PAHs have been detected both in the halo of normal galaxies 
like NGC 891 and in the outflows of starburst galaxies like M82.
Our work shows that in both these environments the lifetime
against CR bombarding of large PAHs ($N_{\rm C} >$ 150-200) is comparable 
to or longer than the circulation timescale between disk and halo and
the starburst lifetime ($\sim$200 Myr and $\sim$ 20 Myr respectively). 
PAHs dispersed in the hot gas filling the
galactic halo and the starburst outflow are rapidly destroyed by
collisions with thermal ions and electrons, but this mechanism is
inefficient if the molecules are isolated from this gas in denser
cloudlets.
CRs can access the denser clouds and together with X-rays will set the
lifetime of those protected PAHs, which can be used as a `dye' for
tracing the presence of cold entrained material.

In cooling flow galaxy clusters like A85 and Virgo the cosmic
ray intensity is remarkably enhanced with respect to the solar
neighborhoods, as a consequence the PAH lifetime is much shorter.
Nevertheless, the survival time against CR bombardment is at least two orders of 
magnitude longer than the PAH lifetime in a hot gas (1 -- 10$^3$ yr), implying
that the molecules will be rapidly destroyed in the gas phase
of the ICM. They could survive if protected in some cold
entrained material and in this case the PAH lifetime will be set by 
CRs. Future observations would hopefully provide
more insights about the validity of our predictions.

The major source of uncertainty in the determination of the time scale
for PAH processing by CRs resides in the choice of the nuclear
threshold energy $T_0$ (for electron collisions) and the fragment binding
energy $E_0$ (for ion collisions).
Our conclusions are robust despite
the large variability in the PAH lifetime 
induced by the incertitude on the above parameters. Nevertheless this 
variation emphasizes again the importance of a better determination of 
these quantities.

We find that thermal ions/electrons in a hot gas are much more
effective in destroying PAHs than CRs. This is due to the fact
that the stopping power of the thermal ions/electrons under consideration
($T \sim$ 10$^7$ K, $n_{\rm e}\sim$ 0.1 cm$^{-3}$) is high, close to its maximum 
value, and allows the transfer into the molecule of enough energy to have
the dissociation probability close to one (the dissociation probability
increases with the transferred energy). Almost any ion/electron is
able to destroy a PAH, and the resulting destruction efficiency is
very high (cf. MJTa and MJTb).

On the other hand, in the energy range we consider here for CR ions (5
MeV/nucleon - 10 GeV), the stopping power, and then the energy
transferred into the PAH and the dissociation probability for PAHs,
decrease rapidly for increasing energy of the incoming ion. For CR
electrons (5 MeV - 10 GeV) the cross section for carbon atom removal
is almost constant but very small ($\sim$ 6$\times$10$^{-23}$ cm$^2$).
For both ions and electrons the CR spectra are decreasing functions of
the energy. The combination of these two factors implies that only few
CRs are able to destroy PAHs, resulting in a destruction efficiency
very low compared to thermal ions/electrons in a hot gas.

Our results show that CRs set a timescale for the destruction 
of PAHs with less than a $\simeq100$ C-atoms of only 100 Million years 
in the interstellar medium. Larger PAHs or very, very small dust grains
 -- with $N_{\rm C} >$ 100 atoms -- are predominantly processed by 
interstellar shock waves (rather than CRs) on a very similar 
timescale ($\tau_{\rm shock}\simeq 150$ Myr; MJTa).

PAH molecules are modelled to be efficiently formed as molecular 
intermediaries or sideproducts of the soot formation process in the 
stellar ejecta from C-rich objects, in particular C-rich AGB stars 
\citep{frenklach89, cherchneff92, cherchneff00a, cherchneff00b, cau02}.
If we assume that AGB stars are the primary source for PAH replenishment 
in the ISM, the injection timescale,  
$\tau_{\rm formation}\simeq 3 \times 10^9$ Myr \citep{jones94} is much
longer than the lifetime estimated above.
Thus, there is a factor of about 50 discrepancy between the injection 
and destruction timescales for PAHs in the ISM. This parallels the 
discrepancy between the injection and destruction timescales for 
interstellar grains \citep{jones94}. As for dust grains, the conclusion 
seems to be inescapable that PAHs  must be reformed rather efficiently 
in the ISM itself. 

As suggested by \citet{jones96}, interstellar PAHs 
may be the fragmentation products of carbonaceous grain collisions in 
shocks. Experiments mimicking the effects of grain-grain collisions 
support the formation of PAHs and fullerenes as well as carbon chains 
and clusters from Hydrogenated Amorphous Carbon (HAC) grains \citep{scott97}. 
This would replenish the PAH population on a similar timescale ($\simeq 100$ Myr) 
as the destruction timescale. With the numbers quoted above, if all of 
the carbon were injected into the ISM in the form of HAC grains, this 
scheme would (almost) be in steady state with 5\% of the elemental C 
in the form of PAHs.  However, this comes at the expense of the rapid 
destruction of carbon grains, as seems to be suggested by the work of 
\citet{serra08}. So, this would have to be fit into a general model 
where gaseous carbon accretes and reacts with carbonaceous (e.g., HAC) 
grains in interstellar clouds \citep[cf., ][]{jones90a, jones90b}.

This growth process is then balanced by grain-grain collisions in strong 
shock waves in the intercloud medium producing PAHs that are then destroyed 
through CR interaction in the diffuse ISM. While such a model could 
work numerically, the basic premise --- rapid accretion and reaction of 
gaseous carbon on grains forming predominantly HAC-like layers -- has not 
been demonstrated in the laboratory. Moreover, observationally, depletion 
studies suggest that carbon does not seem to partake in the rapid exchange 
between gas and solid phases in the ISM \citep{tielens09}. Alternatively, 
rather than interstellar PAHs resulting from chemical processes in stellar 
ejecta -- either directly or through the shattering of carbon grains -- 
these species may be largely formed from smaller carbon molecules through 
chemical processes in the ISM. 

Theoretical studies show that ion-molecule 
reactions can form benzene-like species in the dense and warm environments 
of pre-Planetary Nebulae (such as \object{CRL 618}) and in protoplanetary disks around 
young stars \citep{herpin00, woods05, woods07}. 
However, such special environments are few and far apart and not important 
in a model for the global presence of PAHs in the ISM. Models predicts that 
ion-molecule chemistry in cold dense molecular clouds can produce benzene 
but at peak abundances at early times of 10$^{-9}$. Likely more relevant are 
the steady state abundances at long times and those are much less \citep[$10^{-14}$; ][]
{mcEwan99}. So, the gas phase formation of PAHs in molecular clouds 
does not seem a very promising route. Essentially, the basic building block 
of PAHs (acetylene) is not very abundant and there isn't much free carbon 
around to drive the chemistry since most of it is locked up in CO in molecular 
clouds. In conclusion, the origin of interstellar PAHs is still clouded in 
quite some intriguing mystery.
\smallskip
\smallskip

\begin{acknowledgements}
We would like to thank the referee for careful reading and valuable
comments. We are grateful to F. Galliano for providing us with the
values of $G_0$ in NGC 891 and to H. Leroux for useful discussions on
the physics of radiation damage in carbon materials.  E.R.M. thanks
G. Lavaux for support and technical assistance and acknowledges
financial support by the EARA Training Network (EU grant
MEST-CT-2004-504604).  Studies of interstellar PAHs at Leiden
Observatory are supported by the advanced-ERC grant 246876 on ``The
role of large Polycyclic Aromatic Hydrocarbon molecules in the
Universe''.
\end{acknowledgements}


\end{document}